\pdfoutput=1
\documentclass[12pt]{article}
\usepackage{graphicx,amssymb,lipsum,mathtools,amsmath,amssymb}

\def\ea{\end{array}}

\def\bc{\begin{center}}
\def\ec{\end{center}}
\usepackage{graphicx}
\usepackage{xcolor}
\usepackage{fancyhdr}
\usepackage{ae,aecompl}
\usepackage{epsfig}
\usepackage{mathtools}
\usepackage{ mathrsfs }
\usepackage{ upgreek }
\usepackage{ dsfont }
\usepackage{IEEEtrantools}
\usepackage{verbatim}
\usepackage{float}

\usepackage{amsfonts}
\usepackage{amssymb}
\usepackage{lscape}
\usepackage{amsmath}
\usepackage{textcomp}
\usepackage{txfonts}   %% This creates problem in aligning the Acknowledgement below part. But very necessary to keep alignment of text.
\usepackage{pstricks}
\usepackage{wrapfig}
\usepackage{multirow}
\usepackage{booktabs}
\usepackage{textcomp}
\allowdisplaybreaks
\begin{document}

\title{Prospects for the Mass Ordering (MO) and $\theta_{23}$-Octant sensitivity in LBL experiments:
UNO, DUNE \& NO$\nu$A}
\author{Mandip Singh$^*$ \\
\\
{ \it Department of Physics, Centre of Advanced Study, P. U., Chandigarh, India.}\\
{\it Email:manyphy101@gmail.com}
}

\maketitle

\begin{abstract}
 This article represents quantitative numerical analysis 
 to find the sensitivity for the mass ordering and octant of atmospheric mixing angle 
 $\theta_{23}$ within 3$\sigma$ range of oscillation parameters, in the context of
 three long base line (LBL) accelerator experiments viz. UNO, DUNE and NO$\nu$A. 
In order to include degeneracy arising due to Dirac's $\delta_{CP}$ phase, we vary 
this parameter in $-\pi$ to $+\pi$ range.
                             We find that it is possible to 
investigate mass ordering at and around the first oscillation maxima, while second and higher order 
oscillation maximas are inappropriate for such investigations due to very fast oscillations over the possible 
beam energy spread at these oscillation maximas. The probability sensitivity of 
UNO experiment is almost twice that of NO$\nu$A and DUNE experiments. 

We notice that on the basis of quantitative sensitivity pertaining to the event rate, it is
possible to investigate the mass ordering within all experiments. Conclusively, 
like NO$\nu$A and DUNE experiments, UNO experiment stands as better alternative for  investigating
mass ordering, especially when we need to cross check the results at higher beam energies
and base line lengths.

If hierarchy is best known then we are left with octant and $\delta_{CP}$ degeneracies 
that affects the unambiguous determination of these parameters. In order to detect different 
possible degenerate solutions, we have represented comprehensive study in terms of 
contour plots in the test $\theta_{23} - \delta_{CP}$ plane 
for different representative true values of parameters. We observe that discrete solutions viz. 
{\it wrong octant-right $\delta_{CP}$, wrong octant-wrong $\delta_{CP}$} and {\it right octant-wrong $\delta_{CP}$}
and continuous solutions arising due to submergence of discrete solutions with 
true solution are possible up to 3$\sigma$ level. It is UNO experiment that alone 
have the potential to remove these discrete solutions, while both NO$\nu$A and DUNE experiments 
have very poor tendency to remove these discrete solutions especially near the maximal mixing. 
We find that these discrete solutions can be resolved up to 3$\sigma$ level by
the combined NO$\nu$A+DUNE+UNO data set, at all multiple degenerate solutions 
in the true parameter space considered under the study.
Though replacing half the neutrino run with antineutrino run introduces qualitative advantage
because of their different dependences on $\delta_{CP}$, but due to lower cross
section and reduction in the statistics, addition of antineutrino data make the precision
worse. Thus considering experimental data only in the neutrino mode, 
enhances $\delta_{CP}$ and $\theta_{23}$ precision significantly.

We observe that for synergistically combined data set, the CP precision is seem to be better 
for $\delta_{CP} = \pm 90^0$ as compared to $\delta_{CP} = 0^0$
for a given true value of $\theta_{23}$. While the $\theta_{23}$
precision at given value of $\delta_{CP}$ is worse near the maximal mixing and improves
as one moves away.
%%%%
\end{abstract}

%%%%%%%%%%%%%%%%%%%%%%%%%%%%%%%%%%%%%%%%%%%%%%%%%%%%%%%%%%%%%%%%%%%%%%%%%%%%%%%%%%%%%%%%%%%%%%%%%%%
\section{Introduction and motivations}
\label{section:Introduction}

It is well established that neutrinos are very tiny massive entities and their flavour states are mixed, 
    due to which one flavor state can transform in to another during their time evolution in space and medium. 
     These unique characteristics have emphasized to look beyond 
       standard model (SM), as it's extension to include the non zero mass of 
        neutrinos. Over past few decades, an exciting era 
         of neutrino physics has signed its golden
         foot prints in the form of assigning firm limits on the atmospheric, solar 
          and the reactor mixing angles and tiny mass squared differences 
       via the dedicated neutrino experiments involving neutrinos 
      from sun\cite{sol1}--\cite{sol3}, atmosphere\cite{atm2}, nuclear 
     reactors\cite{nucre1}--\cite{nucre3} 
   and accelerators\cite{acc1}--\cite{acc2}. In order to exclude a large subset of 
   the still broader ranges of the parameter space, we need 
high-precision measurements of the neutrino oscillation parameters, that would provide
crucial hints towards our understanding of the physics of neutrino masses and mixing 
\cite{numix}--\cite{nuoscpara}. 
   
   Among the left unknowns in the physics of neutrino oscillation
   are the leptonic CP-violation phase
  $\delta_{CP}$, the mass ordering (MO) of neutrino mass eigen states 
  and the octant of atmospheric mixing angle $\theta_{23}$. 
  All the three global fits \cite{dformt}--\cite{mgartsto} point that the 
  value of atmospheric mixing angle $\theta_{23}$ deviates from maximal 
  mixing (MM) (i.e. $\theta_{23} \neq 45^0$). A 3$\sigma$ range $36^0 \leq \theta_{23} \leq 54^0$, 
  suggests that its value could be less than or grater than $45^0$.
 In the present work we will especially discuss of the possible investigation of mass ordering (MO)
and octant sensitivity (in the upper octant (UO) $\theta_{23} > 45^{o}$ and the lower octant (LO) $\theta_{23} < 45^{o}$)
of atmospheric mixing angle $\theta_{23}$ in the context of three experimental setups: 
UNO-Henderson [L=2700 Km], DUNE(LBNE) [L=1300 Km] and NO$\nu$A [L=812 Km]. It has been observed
that the main complication that arises in the determination of $\theta_{23}$-octant is due to the unknown value
of $\delta_{CP}$ in the subleading terms of $P_{e \mu}$ channel 
which gives rise to octant-$\delta_{CP}$ degeneracy. It
was discussed in \cite{phubwitr}\cite{hminka} that combining the reactor measurement of $\theta_{13}$
with the accelerator data will be beneficial for the extraction of information on the 
octant value from the $P_{e \mu}$ channel.
Recently, it has been realized that for the appearance channel, the octant degeneracy
can be generalized to the octant-$\delta_{CP}$ degeneracy corresponding to any value of $\theta_{23}$ in the
opposite octant \cite{skraut}, \cite{pclom}. A continuous generalized degeneracy in the three-dimensional
$\theta_{23} - \theta_{13} - \delta_{CP}$ plane has been studied in Ref.\cite{pclom}.

 It is evident from the theoretical formalisms \cite{akhmet}, \cite{jsato} that the 
 confirmation of moderately large value of 
    reactor mixing angle $\theta_{13}$ (in comparison of
       $\theta_{13}=0^{o}$) \cite{int1}--\cite{th13} has increased the possibility of the investigation of 
             mass hierarchy (MH) and has also increased the possibility of non zero value of
        the CP-violation phase $\delta_{CP}$.
        Long Base Line (LBL) neutrino experiments 
  due to their long base lines have advantage over the short base 
   line experiments, latter can be approximated to vacuum oscillations. Over the long 
   distances of experimental base lines, contamination of matter effects enhances the amplitude of  
   the transition probabilities to an extent that they become sensitive to the experiments.
   The CP conjugate of oscillation probability
   can be obtained by merely changing the sign of CP-violation phase $\delta_{CP}$ 
   and matter potential $`A'$, due to which matter effects add to vacuum effects
      (in NH case), which makes transition probability amplitude so large at moderate 
       base line lengths, that we expect them to detect experimentally. But now if
        we shift from normal mass hierarchy (NH i.e. $\Delta m_{13}^{2}>0$)
         to inverted mass hierarchy (IH i.e. $\Delta m_{13}^{2}<0$),
         the mass hierarchy parameter $\alpha$ in Eqn. (\ref{seri})
           changes sign, due to which a part of matter 
            effects get subtracted, which lower the 
            value of probability amplitude.
             This addition in the case of NH 
              and subtraction 
              in the IH case,
              separates
   the NH and IH probabilities
    to an extent that we expect them to differentiate experimentally. 
    
    The determination of $\delta_{CP}$ in long-baseline experiments
is constrained by the parameter degeneracy \cite{degena} -- \cite{degene}. 
In particular, the limited hierarchy and octant sensitivity of these experiments, give
rise to hierarchy-$\delta_{CP}$ degeneracy and
octant-$\delta_{CP}$ degeneracy. The behavior
of hierarchy-$\delta_{CP}$ degeneracy is similar in the neutrino and anti-neutrino
oscillation probability \cite{skraut} but the octant-$\delta_{CP}$ degeneracy behaves
differently in neutrinos and anti-neutrinos \cite{skagwa}. So while determining $\delta_{CP}$ phase,
addition of anti-neutrinos over neutrinos can help
in removing the wrong octant solutions but not the wrong hierarchy solutions. 
Apart from the synergy described above, the
role played by anti-neutrinos also depends on the baseline and
flux profile of a particular experiment \cite{mghosh}. 
The current best-fit value for $\delta_{CP}$ is close to $- \pi/2$, although
at $3 \sigma$ C.L. the whole range of [0, 2$\pi$] remains allowed \cite{glbfit1}, \cite{glbfit2}.

    At long base lines the mass hierarchy asymmetry in the oscillation probabilities is larger
     than the CP violation effect arising due to the variation of the $\delta_{CP}$ phase
     over the full range (i.e. $-\pi ~to ~ + \pi$),
      which makes these suitable to determine the mass hierarchy 
        as well as to constrain the $\delta_{CP}$ phase\cite{mocpv}.  
        The recently proposed neutrino oscillation experiment DUNE \cite{dune1}, 
     has baseline nearly equal to the previously proposed LBNE experiment. 
     The possibility of measuring the neutrino
mass hierarchy and octant sensitivity 
in atmospheric and reactor neutrino experiments 
have been considered in details in the literature 
\cite{atmex1}--\cite{atmex18}. In \cite{atmex12}, the octant--$\theta_{13}$, 
octant--$\delta_{CP}$ and intrinsic octant degeneracies and their possible 
resolution has been discussed. 
The octant degeneracy is different for neutrinos and antineutrinos and hence a
combination of these two data sets can be conducive for the removal of this degeneracy for
most values of $\delta_{CP}$ \cite{skagwa}, \cite{degenc} \cite{mgskgos}.
It has been concluded in \cite{ntnath_mg}, that 
for base line L=1300 Km (DUNE), a run time of 10 years in the neutrino 
mode only is appropriate to have observable $\delta_{CP}$ sensitivity over 
the entire [-$\pi$, +$\pi$] range and run time of [7,3] \& [5,5] in 
[neutrino, anti-neutrino] mode is optimal for observable $\theta_{23}$ octant 
sensitivity. 

In particular many papers have discussed possibilities of the resolution of the degeneracies discussed in 
above paragraphs by using different detectors in the same experiment \cite{detc1}-\cite{detc3}. The synergistic combination of data
from different experiments was also discussed as an effective means of removing such 
degeneracies by virtue of the fact that the oscillation probabilities offer different combinations of
parameters at varying baselines and energies \cite{exp1}-\cite{exp9}. In \cite{atmex17} it has been shown that with the high
precision measurement of $\theta_{13}$ by reactor experiments, the degeneracies can be discussed in
an integrated manner in terms of a generalized hierarchy - $\theta_{23}$ - $\delta_{CP}$ degeneracy.

In the present work we will consider the case of [10,0] mode, as
neutrino ($\nu_e \rightarrow \nu_\mu$ channel) produced in the decay of 
$\mu^+$ mesons in the accelerator experiments, as can be seen in Eqn.~\ref{mupm}.
We will show that merely the addition of UNO experiment data to the NO$\nu$A and DUNE experiments
for the 10 years of run time without caring of anti-neutrino oscillation mode is helpful to remove 
octant - $\delta_{CP}$ degeneracy for known hierarchy. The known hierarchy is chosen as NH and beam 
as the on axis.
       Though the combined capability of NO$\nu$A experiment with T2K, DUNE and ICAL etc. 
       experiments in hierarchy, octant and
$\delta_{CP}$ determination has been investigated in the literature so far, 
but a comprehensive study of the removal of degeneracies
using these three facilities (i.e.NO$\nu$A, DUNE, UNO) together is an unique one.

We also present the precision of the parameters $\theta_{23}$ and $\delta_{CP}$ from the combined
analysis of data from NO$\nu$A, DUNE, UNO experiments. Though there is a qualitative advantage of including both neutrino and
antineutrino channels because of their different dependences on $\delta_{CP}$, this advantage is squandered by
the lower cross section of antineutrinos. Also, replacing half the neutrino run with antineutrinos 
reduces the statistics and hence the precision becomes worse. 
Rather, running in the neutrino mode gives enhanced statistics and hence better precision.

%%%%%%%%%%%%%%%%%%%%%%%%%%%%%%%%%%%%%%%%%%%%%%%%%%%%%%%%%%%%%%%%%%%%%%%
\section{Theoretical methodology} 
\label{section:phenomenology} 
%%%%
Decay of $\mu^+$ and $\mu^-$ mesons in to long tunnels can be given as
\begin{eqnarray*}
 \mu^+ \longrightarrow e^+ ~ + ~\nu_e ~ + ~ \overline{\nu}_\mu   \\
  \mu^- \longrightarrow e^- ~ + ~ \overline{\nu}_e ~ + ~ \nu_\mu
  \label{mupm}
\end{eqnarray*}
hence the different possible flavour channels that can be studied with $\mu^+$ mesons are 
\begin{eqnarray*}
 \nu_e &\longrightarrow& \nu_\mu ~; ~~~~~ \overline{\nu}_\mu \longrightarrow \overline{\nu}_\mu  \\
  \nu_e &\longrightarrow& \nu_e ~; ~~~~ \overline{\nu}_\mu \longrightarrow \overline{\nu}_e
\end{eqnarray*}
and the different possible flavour channels that can be studied with $\mu^-$ mesons are 
\begin{eqnarray*}
  \overline{\nu}_e &\longrightarrow& \overline{\nu}_\mu ~; ~~~~~ \nu_\mu \longrightarrow \nu_\mu  \\
 \overline{\nu}_e &\longrightarrow& \overline{\nu}_e ~; ~~~~ \nu_\mu \longrightarrow \nu_e
\end{eqnarray*}
%%%
The analytic expressions for the neutrino flavor transition probabilities up to first and/or 
  second order of small parameters viz. the mass ordering parameter $\alpha$ 
  and third mixing angle `$\theta_{13}$' 
    also known as reactor mixing angle, has been calculated in the literature by \cite{hminkta}, \cite{acrvera},
      \cite{amgago} and \cite{akhmet} very throughly. These analytic expressions hold very well within certain limits of
       baseline $`L'$ and beam energy $`E'$. The transition probability of oscillation for the golden channel 
       in case of particle and anti-particle channels can be written as
%%%          
 \begin{eqnarray}
    P_{e \mu}^\pm &=& P_a ~ + ~ 4 s_{13}^2 s_{23}^2 (Y^\pm)^2     
                     ~  +  ~2 ~\alpha~ s_{13} ~sin {2 \theta_{12}} ~sin {2\theta_{23}}~
                      cos{\left(\Delta \frac{L}{2} \mp \delta_{CP}\right)}
       \label{seri}
         \end{eqnarray}
%%%         
where 
\begin{eqnarray*}
  P_a &=& \alpha^{2} ~sin^{2} {2\theta_{12}}~ c^{2}_{23}~ 
    X^2   \nonumber \\
\label{coeff}
\end{eqnarray*}
%%%%
Also, we can write the probability expression for the $\nu_{\mu}$ disappearance channel as
\begin{eqnarray}
 P_{\mu \mu}^{\pm} &=& 1 - sin^{2} 2 \theta_{23} ~ sin^{2} ~ \left[\Delta \frac{L}{2} \right] + \frac{\upalpha ~ \Delta ~ L}{2} ~ c_{12}^{2} 
~ sin^{2} 2 \theta_{23}  ~ sin ~ \Delta L       \nonumber  \\
& & - ~ \alpha^{2} sin^{2} 2 \theta_{12} ~ c_{23}^{2} ~X^2 
 - \left(\frac{\alpha ~ \Delta ~ L}{2}\right)^{2} c_{12}^{4} ~ sin^{2} 2 \theta_{23} ~ cos ~\Delta L \nonumber  \\
 & & \pm ~ \frac{\alpha^{2}}{2 A} ~ sin^{2}2 \theta_{12} ~ sin^{2}2 \theta_{23} ~\left( sin \left[ \Delta \frac{L}{2} \right]~ cos \left[(A \mp 1) \Delta \frac{L}{2}\right] ~ X 
 - \frac{\Delta ~L}{4} ~sin ~ \Delta L \right)  \nonumber   \\
 & & - ~ 4 ~ s_{13}^{2} ~ s_{23}^{2} (Y^{\pm})^2     \nonumber    \\ 
 & & \mp ~ \frac{2}{A \mp 1} ~ s_{13}^{2} ~ sin^{2} 2 \theta_{23} \left( sin\left[\Delta \frac{L}{2}\right] ~ cos \left[A \Delta \frac{L}{2}\right] 
 Y^{\pm} \mp \frac{A~ \Delta ~L}{4}~ sin ~ \Delta L \right)  \nonumber   \\
 & & - ~ 2 ~ \alpha ~ s_{13} ~sin ~2 \theta_{12} ~sin ~2 \theta_{23} ~ cos~\delta_{CP} ~cos\left[\Delta \frac{L}{2}\right] ~X ~Y^{\pm}   \nonumber  \\
 & & \pm ~ \frac{2}{A \mp 1}\alpha ~ s_{13} ~ sin ~2 \theta_{12}~ sin ~2 \theta_{23} ~ cos ~2 \theta_{23}~ cos~\delta_{CP}~ sin\left[\Delta \frac{L}{2}\right] \nonumber  \\
 & & \times  \left( \pm A ~ sin\left[\Delta \frac{L}{2} \right] - cos \left[(A \mp 1)\Delta \frac{L}{2}\right] ~X
 \right) \nonumber  \\
 \label{pmm}
\end{eqnarray}
%%%%
with
%%%%
\begin{eqnarray}
X =  \frac{sin~{[A \Delta \frac{L}{2}]}}{A} ~;   ~~~~~~~
Y^{\pm} = \frac{sin{[(A \mp 1) \Delta} \frac{L}{2}]}{(A \mp 1)}
\label{srcoeff}
\end{eqnarray}
%%%%%%
where upper sign corresponds to particle probability case and lower sign to 
the anti-particle case. Anti-particle probability can be obtained from that of 
particle case by merely changing  $\delta_{cp} \rightarrow -\delta_{cp}$ and $V \rightarrow -V$(or $A\rightarrow-A$).

Here $A \equiv 2~E~V/\Delta m_{31}^{2}$, where $V = \sqrt{2} ~G_{F} ~N_{e}$; with $N_{e}$ 
  is the number density of the electrons in the medium; $G_{F}$ = Fermi weak coupling 
    constant = $11.6639 \times 10^{-24} ~eV^{-2}$, $\Delta \equiv
     \Delta m_{31}^{2}/2~E \simeq \Delta m_{32}^{2}/2~ E$,
      $\alpha = \Delta m^{2}_{21}/\Delta m^{2}_{32}$.                  
%
%
%%%%%%%%%%%%%%%%%%%% Table  %%%%%%%%%%%%%%%%%%%%%%%%%%%%%%%%%%
\setlength{\arrayrulewidth}{0.27 mm}  %% controlls line width 
\setlength{\tabcolsep}{11pt}  
\renewcommand{\arraystretch}{1.5} 
\newcommand{\head}[1]{\textnormal{\textbf{#1}}} 
%%%%%%%%%%%%%%%%%%%%%%%%%%%%%%%%%%%%%%%%%%%%%%
\begin{table*}[htbp!]
\centering
\caption{The best fit and 3$\sigma$ values of mixing angles and mass square differences
from global fit of neutrino oscillation data, adopted from \cite{glbft}, \cite{glbgarcia}.
\label{tab:bfit}}
\begin{tabular}{ p{3cm} p{3cm} p{3cm} }  \hline
 \head{Parameter} &   \head{best fit} {{\bf \fontsize{11}{31}\selectfont$\pm 1\sigma$}} 
 & {{\bf \fontsize{11}{31}\selectfont 3$\sigma$}}  \\ 
%%%& & & exact & approximate \\ 
\hline
$\theta_{12}^o$ & $34.6 \pm 1.0$ & 31.29 -- 37.8  \vspace{0.2cm}\\ 
$\theta_{23}^o$[NH] & $42.3^{+3.0}_{-1.6}$ & 38.2 -- 53.3  \\
$\theta_{23}^o$[IH] & $49.5^{+1.5}_{-2.2}$ & 38.6 -- 53.3  \vspace{0.2cm}\\
$\theta_{13}^o$[NH] &  $8.8 \pm 0.4$ &  7.7 -- 9.9  \\
$\theta_{13}^o$[IH] &  $8.9 \pm 0.4$ &  7.8 -- 9.9  \vspace{0.2cm}\\
\fontsize{19}{31}\selectfont {$\frac{\Delta m_{21}^2}{10^{-5} ~eV^2}$} & $7.6^{+0.19}_{-0.17}$ &  7.02 - 8.18  \vspace{0.3cm}\\  
\fontsize{19}{31}\selectfont {$\frac{|\Delta m_{31}^2|_{NH}}{10^{-3} ~eV^2}$} & $2.48^{+0.05}_{−0.07}$  &  2.30 -- 2.65  \vspace{0.3cm}\\
\fontsize{19}{31}\selectfont {$\frac{|\Delta m_{31}^2|_{IH}}{10^{-3} ~eV^2}$} & $2.44^{+0.048}_{-0.047}$  &  2.20 -- 2.59  \vspace{0.1cm}\\
\hline
\end{tabular}
%%%
\end{table*}
%     
%%
%  
%%%
%%%%%%%%%%%%%%%%%%%%%%%%%%%%%%%%%%%%%  Figure here  %%%%%%%%%%%%%%%%%%%%%%%%%%%%%%%%%%%%%%%%%%%%%%%%%%%%%%%%%%
\begin{figure*}[h!]
\centering
\caption{(Color online) Oscillogram of probability in the beam energy `E' and baseline length `L' plane.
          Probability scale in the legend bar on RHS is in \%. The diagonal green lines show 
          first and second oscillation maxima (O.M.).}
\includegraphics[width=0.8\textwidth]{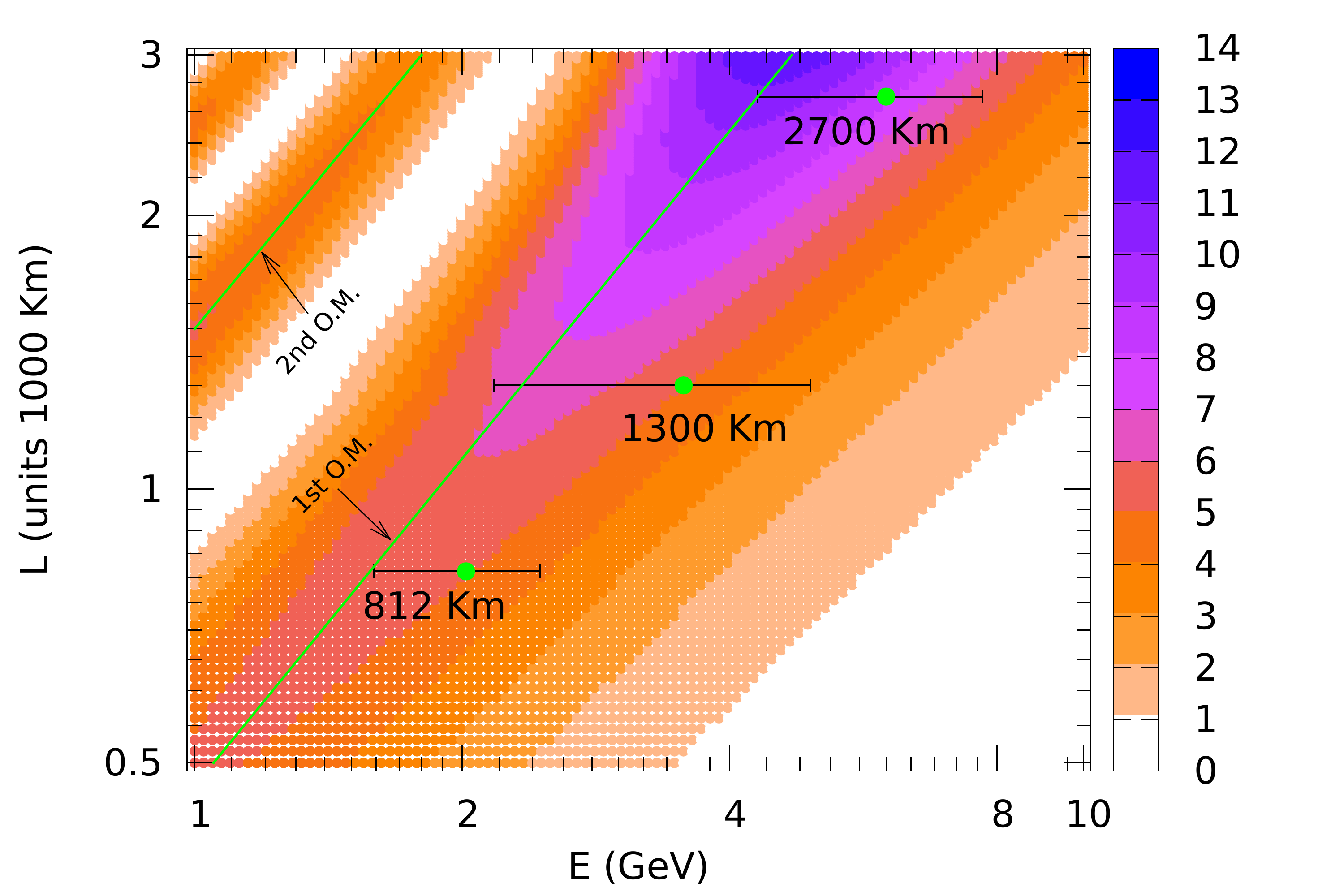}
\label{fig:prb_el}
\end{figure*} 
%  
%%%

The differential event rate for given base line length `L' and muon energy $E_\mu$ for 
particular channel `i' can be written as \cite{frudhub}
\begin{eqnarray}
 \frac{dn_i}{dE} = \left[ N_{\mu^i} ~N_{KT} ~\varepsilon ~\frac{10^9 ~N_A}{m_\mu^2 ~\pi} \right] 
      \left[ \frac{E_\mu}{L^2} ~f_i(E, E_\mu) ~\sigma_i(E) \right] \left[ P_i (E)\right] ~T
      \label{diffent}
\end{eqnarray}
%%%
where first square bracket represents the normalization factor, second to the flux component at the detector site 
and third square parentheses to the oscillation probability and `T' is the run time of the experiment.
$N_{\mu^i}$ is the number of decays of $\mu^\pm$ mesons per year ($N_{\mu^\pm} = 3 \times 10^{20} decays/Year$), $N_{KT}$ is the 
detector mass in kilo tons, $\varepsilon$ is the detector efficiency, $N_A$ is the Avogadro's number, $m_\mu$ ($ = 0.106 ~GeV/c^2$) 
is the muon mass, $E_\mu$ is the muon energy, E is the neutrino energy. The normalized initial spectrum of $\nu$'s 
produced in the decay of unpolarized muons can be written as 
\begin{eqnarray}
 f(E, E_\mu) &\equiv& g_{\nu_e} = g_{\overline{\nu}_e} = 12 ~\frac{E^2}{E_\mu^3} (E_\mu -E)  \nonumber \\
  &\equiv& g_{\nu_\mu} = g_{\overline{\nu}_\mu} = 2 ~\frac{E^2}{E_\mu^3} (3 ~E_\mu - 2 ~E) 
  \label{nuspect}
\end{eqnarray}
Now the charged current neutrino cross sections per nucleon in the detector for neutrino of energy `E' can be written as
\begin{eqnarray}
 \sigma_i(E) &\equiv& \sigma_{\nu_\mu} (E) = \sigma_{\nu_e} (E) = 0.67 \times 10^{-38} ~\frac{E}{GeV} ~~cm^2  \nonumber \\
             &\equiv& \sigma_{\overline{\nu}_\mu} (E) = \sigma_{\overline{\nu}_e} (E) = 0.34 \times 10^{-38} ~\frac{E}{GeV} ~~cm^2
             \label{xsec}
\end{eqnarray}
%%%
We can find the number of events generally as
\begin{eqnarray}
 N = \sum_i \left( \frac{dn}{dE} \right)_{E=E_i} ~ \Delta E
 \label{entnum}
\end{eqnarray}
where the subscript `i' runs over the number of energy bins.
In the limit $\Delta E \longrightarrow 0$, we have
\begin{eqnarray}
 N = \int_{E_{min}}^{E_{max}} \frac{dn}{dE}  ~ dE
 \label{enttot}
\end{eqnarray}
%%%%%%%%%%%%%%%%%%%%%                                                                                                  %%%%%%%%%%%%%%%%%%%%%%%%%
%%%%%%%%%%%%%%%%%%%%%%%%%%%%%%%%%%%%%%%%%%%%%%%%%%%%%%%%%%%%%%%%%%%%%%%%%%%%%%%%%%%%%%%%%%%%%%%%%%%%%%%%%%%%%%%%%%%%%%%%%%%%%%%%%%%%%%%%%%%%%%%%  
%%%%%%%%%%%%%%%%%%%%%%%%%%%%%%%%%%%%%%%%%%%%%%%%%%%%%%%%%%%%%%%%%%%%%%%%%%%%%%%%%%%%%%%%%%%%%%%%%%%%%%%%%%%%%%%%%%%%%%%%%%%%%%%%%%%%%%%%%%%%%%%%
%%%%%%%%%%%%%%%%%%%%%%                                                                                                %%%%%%%%%%%%%%%%%%%%%%%%%%
                      \section{ Mass ordering (MO) sensitivity parameter 
                                      {{\bf \fontsize{15}{25}\selectfont $A^h$ }}}
                                  \label{sec:mooptimization}
%%%%%%%%%%%%%%%%%%%%%                                                                                                  %%%%%%%%%%%%%%%%%%%%%%%%%
%%%%%%%%%%%%%%%%%%%%%%%%%%%%%%%%%%%%%%%%%%%%%%%%%%%%%%%%%%%%%%%%%%%%%%%%%%%%%%%%%%%%%%%%%%%%%%%%%%%%%%%%%%%%%%%%%%%%%%%%%%%%%%%%%%%%%%%%%%%%%%%%

It is evident from Eqns.~(\ref{seri}), (\ref{pmm}) and (\ref{srcoeff}), transition probability 
in case of IH can be written by replacing $\alpha \rightarrow -\alpha$, $A \rightarrow -A $ 
and $\Delta \rightarrow -\Delta$, as
%%   see p.36-37 of R-XIV
\begin{eqnarray}
 P_{e \mu}^{IH} & \equiv & P_{e \mu}^{NH}(\alpha \rightarrow -\alpha, A \rightarrow -A,  
 \Delta \rightarrow -\Delta )  \nonumber   \\
 & = & P_a + 4 ~s_{13}^2 ~s_{23}^2 ~(Y^-)^2 - 2 ~\alpha ~s_{13} ~sin ~2 \theta_{12}  
       ~sin ~2 \theta_{23} ~cos (\Delta L/2 + \delta_{CP}) ~X ~Y^- 
\label{probih}
\end{eqnarray}
%%% 
In the above equation, for writing convenience, the +ve sign referring to the particle case has been dropped. 

Now we can define a new parameter $A_{e \mu}^h$ as
\begin{eqnarray}
 A_{e \mu}^h & = & P_{e \mu}^{NH} - P_{e \mu}^{IH}  \nonumber  \\
             & = & 4 ~s_{13}^2 ~s_{23}^2 \left( {Y^+}^2 - {Y^-}^2\right) 
             + 2 ~\alpha ~s_{13} ~sin ~2 \theta_{12} ~sin ~2 \theta_{23} \times \nonumber \\
           && \times \left[ \left( cos \frac{\Delta L}{2} ~cos ~\delta_{CP} + sin \frac{\Delta L}{2} ~sin ~\delta_{CP} \right) ~Y^+ 
             + \left( cos \frac{\Delta L}{2} ~cos ~\delta_{CP} - sin \frac{\Delta L}{2} ~sin ~\delta_{CP} \right) ~Y^-  \right] ~X
             \nonumber  \\
 \label{moahpra}
\end{eqnarray}
%%%%
which can be solved further to the following form 
\begin{eqnarray}
 A^h_{e \mu} = 4 ~s_{13}^2 ~s_{23}^2 \left( {Y^+}^2 - {Y^-}^2\right) + cos \left( \beta - \delta_{CP} \right)
 \label{ahemuf}
\end{eqnarray}
with 
\begin{eqnarray*}
  \beta = tan^{-1} \left[ tan \left(\frac{\Delta L}{2}\right) \frac{Y^+ - Y^-}{Y^+ + Y^-} \right]
\end{eqnarray*}
A similar type of expression can be obtained for $\nu_\mu$ disappearance  channel i.e. $A_{\mu \mu}^h$.
%  
%%%
%%%%%%%%%%%%%%%%%%%%%%%%%%%%%%%%%%%%%  Figure here  %%%%%%%%%%%%%%%%%%%%%%%%%%%%%%%%%%%%%%%%%%%%%%%%%%%%%%%%%%
\begin{figure*}[htbp!]
\centering
\caption{(Color online). For LHS column (L=2,700 Km) $\rho_{avg} = 3.8 ~gm/cm^{3}$; for
 middle column (L=1,300 Km) $\rho_{avg} = 3.5 ~gm/cm^{3}$ and for RHS column (L=8,12 Km)
 $\rho = 2.8 ~gm/cm^{3}$. The yellow colored curve shows the event rate for the NH-case, while green curve 
  that in the IH-case and red curve shows the difference between yellow and green curves, `$N_{e ~\mu}^h$'.
 Top row of sub-figures corresponds to $\delta_{CP} = 90^0$ and below rows to 
 $\delta_{CP} = 60^0, ~45^0, ~30^0, ~0^0$ respectively. We choose $E_\mu = 50$ GeV, $N_{\mu}^+ = 3 \times 10^{20}$ 
 muon events per year, $N_{KT} = 10$ KT, $T= 1$ Year.
  All the other vacuum oscillation parameters have been chosen as the 
   best fit values as in table \ref{tab:bfit}.}
\includegraphics[width=0.99\textwidth]{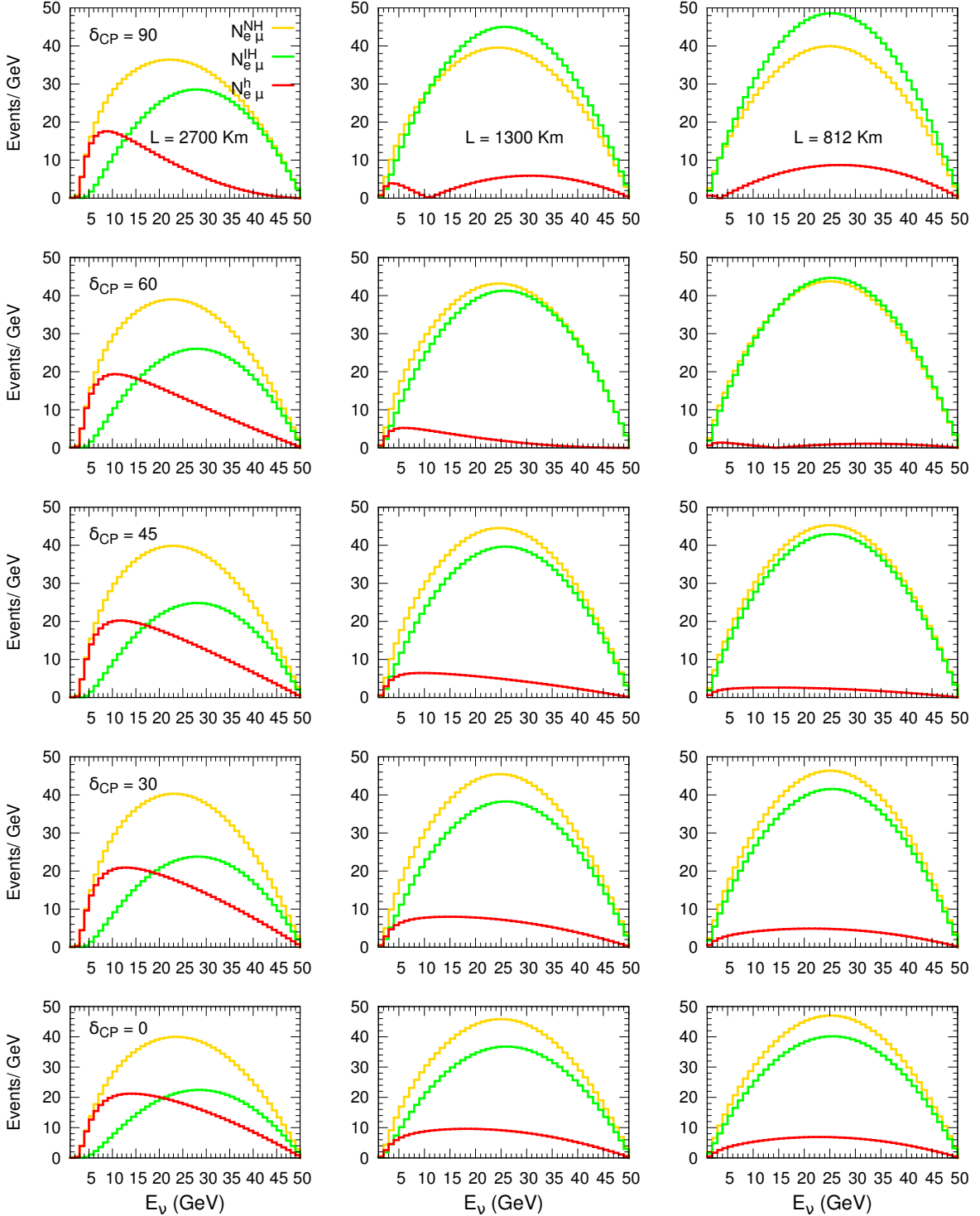}
\label{fig:10kty}
\end{figure*} 
%  
%%%
%%%%%%%%%%%%%%%%%%%%%%%%%%%%%%%%%%%%%  Figure here  %%%%%%%%%%%%%%%%%%%%%%%%%%%%%%%%%%%%%%%%%%%%%%%%%%%%%%%%%%
\begin{figure*}[htbp!]
\centering
\caption{(Color online). Here we choose $N_{KT} = 50$ KT, $T= 10$ Years. Y-axis scale 
 is in $10^2$ units. Rest of particulars are same as in above Fig.~\ref{fig:10kty}.}
\includegraphics[width=0.97\textwidth]{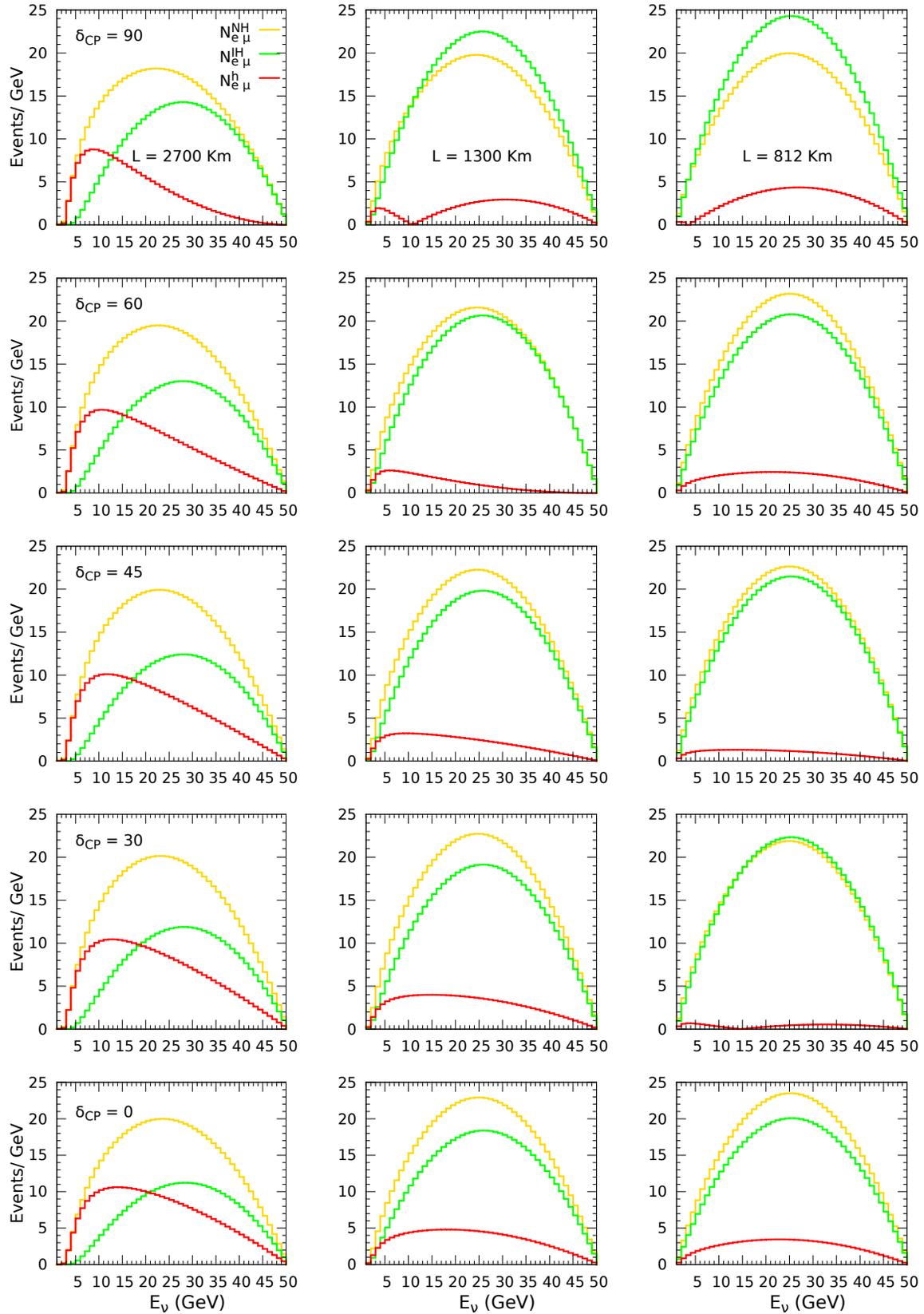}
\label{fig:5hkty}
\end{figure*} 

Now with the help of Eqn's. (\ref{diffent}) to (\ref{enttot}) and above equations, we can write mass hierarchy
parameter in terms of event rate, generally as
\begin{eqnarray}
 N_{\alpha \beta}^h = N_{\alpha \beta}^{NH}  -  N_{\alpha \beta}^{IH} ~
                  \propto ~f_\alpha(E, E_\mu) ~\sigma_\beta (E) ~A_{\alpha \beta}^h ~T
 \label{nabh}                                                                              
\end{eqnarray}
%%%
In the case, we consider the decay of $\mu^+$ mesons, then for the $\nu_\mu$ neutrino flavor (which involves 
$\nu_e \longrightarrow \nu_\mu$ and $\overline{\nu}_\mu \longrightarrow \overline{\nu}_\mu$ channels), the expected 
mass hierarchy sensitivity with the help of Eqn's. (\ref{ahemuf}) and (\ref{nabh}) can be written as
\begin{eqnarray}
 N_{e ~\mu + \overline{\mu}  ~\overline{\mu}}^h = \left[ N^{NH}  -  N^{IH} \right]_{(e ~\mu + \overline{\mu}  ~\overline{\mu})} ~
                  \propto ~ \left( f_{\nu_e}(E, E_\mu) ~\sigma_{\nu_\mu}(E) ~A_{e ~\mu}^h
                           + f_{\overline{\nu}_\mu}(E, E_\mu) ~\sigma_{\overline{\nu}_\mu}(E) 
                               ~A_{\overline{\mu} ~\overline{\mu}}^h \right) ~T
   \label{massentot}
\end{eqnarray}
Sensitivity of $\overline{\nu}_\mu \longrightarrow \overline{\nu}_\mu$ channel towards the investigation of 
mass hierarchy is very low, hence we would not like to study it and also it's combination with $\nu_e \longrightarrow \nu_\mu$ 
channel. We can distinguish among the lepton charges produced as a result of the charged 
interaction of $\nu_\mu$ and $\overline{\nu}_\mu$ in the detector by the application of magnetic field.

We can define an another parameter, the ``mass ordering asymmetry parameter, $A^{asy}$'' in order to get the 
strength of the sensitivity towards the mass ordering investigation with respect to the signal strength, generally as
\begin{eqnarray}
 N^{asy} = \frac{N_{\alpha \beta}^{NH}  -  N_{\alpha \beta}^{IH}}{N_{\alpha \beta}^{NH}  +  N_{\alpha \beta}^{IH}}
 \label{mhasy}
\end{eqnarray}

In Figs.~\ref{fig:10kty} and \ref{fig:5hkty} for the detector exposure of 10 KTY (Kilo Ton Years) and 500 KTY respectively,
we have illustrated the neutrino spectrum at the detector site after traveling the long distance through Earth matter from 
the point of its generation. We have chosen $\mu^+$'s accelerated up to an energy, $E_\mu = 50 ~GeV$. 
With the detector detection threshold neutrino energy of 1 GeV, we divide the neutrino energy spectrum 
in the $E_\nu$ range of (1 - 50) GeV over 49 energy bins, each of size 1~GeV. At $L=2700$ Km event rate is nearly 
independent of the $\delta_{CP}$ phase variations, as is evident from the nearly equal areas under the given colored curve.
In the remaining experiments ($L=1300$ Km \& $L=812$ Km), event rate also feebly depends on the $\delta_{CP}$ phase.
But, as one moves from one base line to another (especially for UNO $\leftrightarrow$ DUNE, or UNO $\leftrightarrow$ NO$\nu$A)
their is observable change in the event rate at given value of $\delta_{CP}$ phase. It is also evident from the second and 
third columns that both DUNE and NO$\nu$A experiments have nearly equal value of total events 
for given type of event rate (i.e. $N_{e ~\mu}^{NH}, ~N_{e ~\mu}^{IH}, ~N_{e ~\mu}^{h}$). Now if we compare 
figures \ref{fig:10kty} and \ref{fig:5hkty}, the event rate at detector configuration 500 KTY is about 
50 times that at 10 KTY.
%%%%
\begin{figure*}[htbp!]
\centering
\caption{(Color online). PREM and line average of Earth's density picked from~\cite{skaggarwal_avgden}.}
\includegraphics[width=0.97\textwidth]{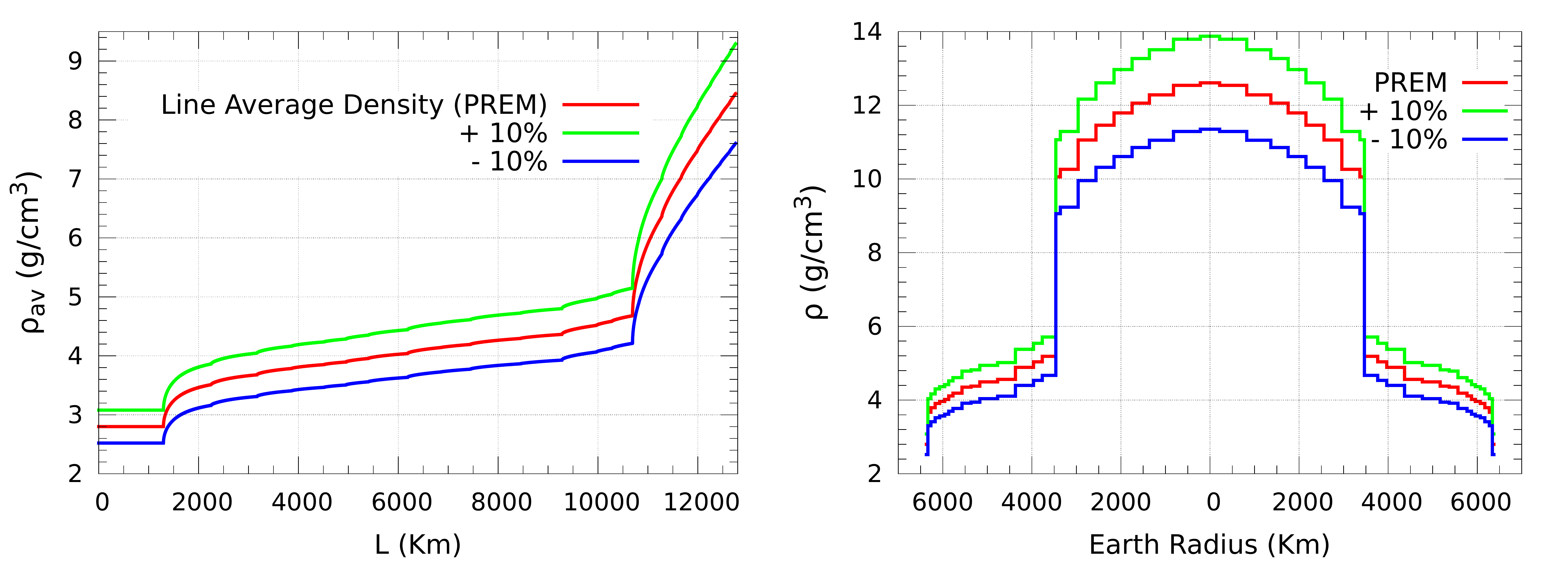}
\label{fig:avgden}
\end{figure*}   
 %%%%
%%%%%%%%%%%%%%%%%%%%%%%%%%%%%% figure %%%%%%%%%%%%%%%%%%%%%%%%%%%%%%%%%
\begin{figure*}[htbp!]
\centering
\caption{(Color online). LHS oscillogram corresponds to the $N_{e ~\mu}^h$ parameter from Eqn.~(\ref{nabh}), 
where event rate is in $10^3$ units, 
RHS oscillogram shows the mass-hierarchy asymmetry parameter $N^{asy}$ from Eqn.~(\ref{mhasy}).
We choose $N_{\mu}^+ = 3 \times 10^{20}$, $E_{\mu} = 50$ GeV, $N_{KT} = 50 ~KT$, $T= 10 ~Y$. 
The value of the terrestrial matter density $\rho_{av} ~in~(gm/cm^3)$ is the average Earth matter
density corresponding to the base line length `L', as shown in Fig.~\ref{fig:avgden}.}
\includegraphics[width=0.97\textwidth]{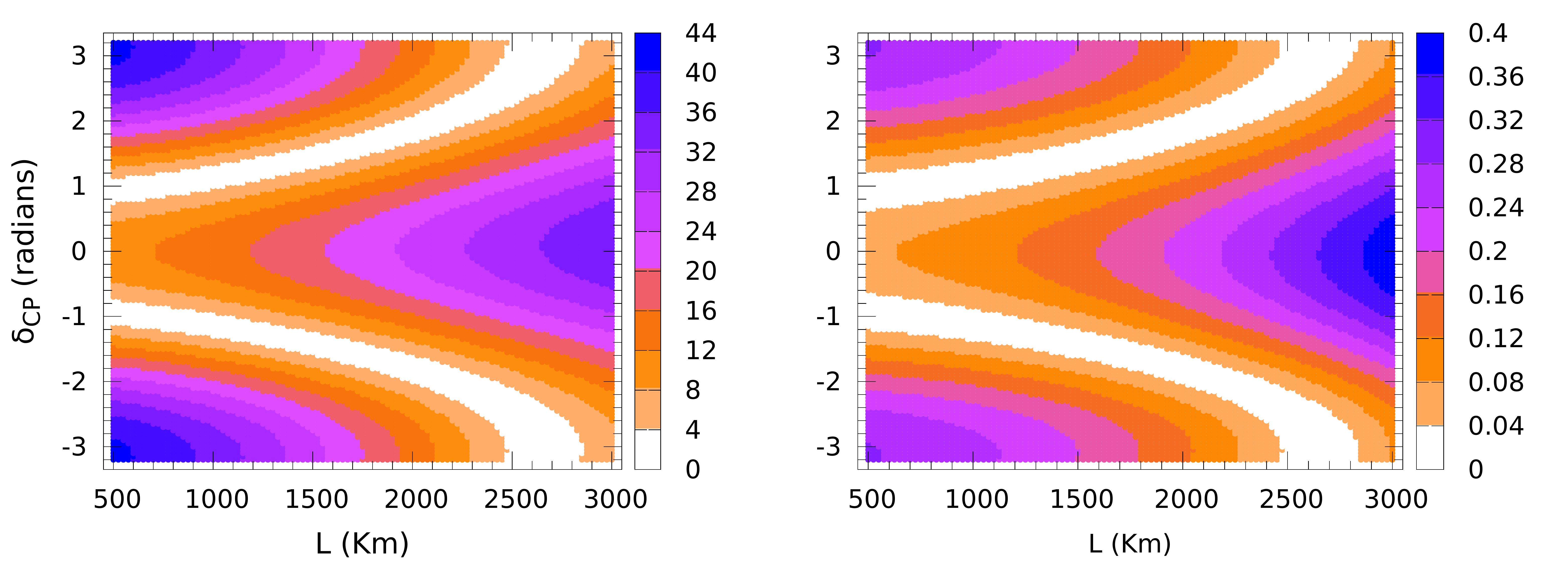}
\label{fig:entlp}
\end{figure*}    

In Fig.~\ref{fig:entlp}, we have illustrated the oscillogram for the mass hierarchy sensitivity 
parameter $N_{e ~\mu}^h$ and mass hierarchy asymmetry parameter $N_{e ~\mu}^{asy}$ in the 
base line `L' and $\delta_{CP}$ plane. We observe that the sensitivity towards the $\delta_{CP}$ phase
is highest in the base line range of $500 \leq L \leq 1000$ Km and also the sensitivity is 
high in the $2500 \leq L \leq 3000$ Km range. But in case of mass ordering asymmetry $N_{e ~\mu}^{asy}$,
the converse is true. In this case asymmetry has highest value in the base line $2500 \leq L \leq 3000$ Km range
and has high value in the $500 \leq L \leq 1000$ Km range. Hence if at short base line range $500 \leq L \leq 1000$ Km,
the $\delta_{CP}$ sensitivity is highest then at long base line range $2500 \leq L \leq 3000$ Km 
mass ordering asymmetry i.e. mass hierarchy sensitivity to the signal ratio is highest. Thus we can conclude to 
say that both NO$\nu$A and UNO experiments represent almost equal suitability to investigate the
$\delta_{CP}$ phase.
%%%
%%%
%%%%%%%%%%%%%%%%%%%%%%%%%%%%%%%%%%% Figure %%%%%%%%%%%%%%%%%%%%%%%%%%%%%%%%%%%%%%%%%%%%%%%%%%%%%%%%%%%%%%%%%%%
\begin{figure*}[htbp!]
%\hspace*{\fill}%
\centering
\caption{(Color online). On LHS is the $N_{e ~\mu}^h$ event rate in $10^3$ units, 
RHS shows mass hierarchy asymmetry parameter $N^{asy}$ for the $\nu_e \longrightarrow \nu_\mu$ channel defined in 
Eqn. (\ref{mhasy}). Here L=2700 Km, $\rho = 3.8 ~gm/cm^3$, $N_{\mu}^+ = 3 \times 10^{20}$, $N_{KT} = 50 ~KT$, $T= 10 ~Y$. }
\hspace*{\fill}%
\includegraphics[width=0.97\textwidth]{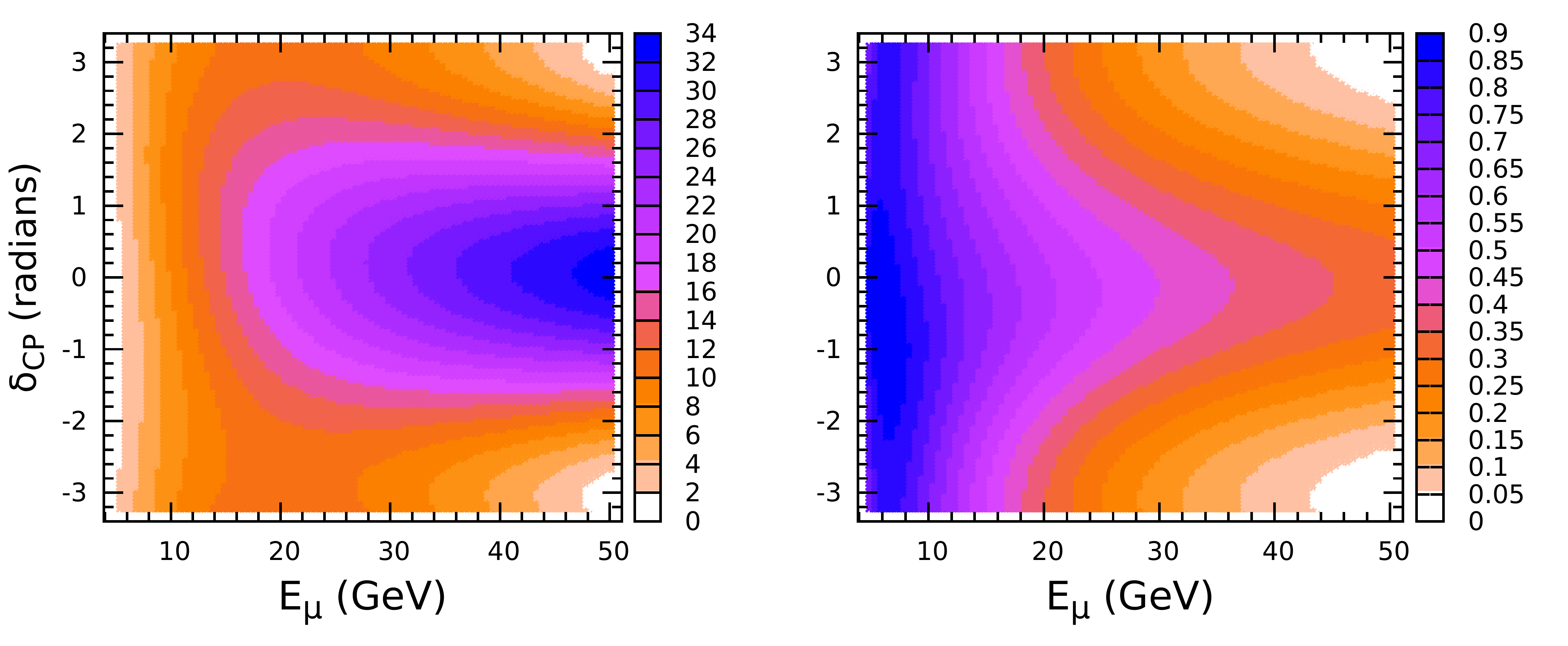}
\label{fig:emu_dcp27}
%\hspace*{\fill}%
\end{figure*}
%%%%%%%%%%%%%%%%%%%%%%%%%%%%%%%%%%%%%%%%%%%%%%%%%%%%%%%%%%%%%%%%%%%%%%%%%%%%%%%%%%%%%%%%%%%%%%%                    
%%%%%%%%%%%%%%%%%%%%%%%%%%%%%%%%%%% Figure %%%%%%%%%%%%%%%%%%%%%%%%%%%%%%%%%%%%%%%%%%%%%%%%%%%%%%%%%%%%%%%%%%%
\begin{figure*}[h!]
%\hspace*{\fill}%
\centering
\caption{(Color online). On LHS is the $N_{e ~\mu}^h$ event rate in $10^3$ units, 
RHS shows mass hierarchy asymmetry parameter $N^{asy}$ for the $\nu_e \longrightarrow \nu_\mu$ channel defined in 
Eqn.~(\ref{mhasy}). Here L=1300 Km, $\rho = 3.5 ~gm/cm^3$, 
$N_{\mu}^+ = 3 \times 10^{20}$, $N_{KT} = 50 ~KT$, $T= 10 ~Y$.  }
\includegraphics[width=0.97\textwidth]{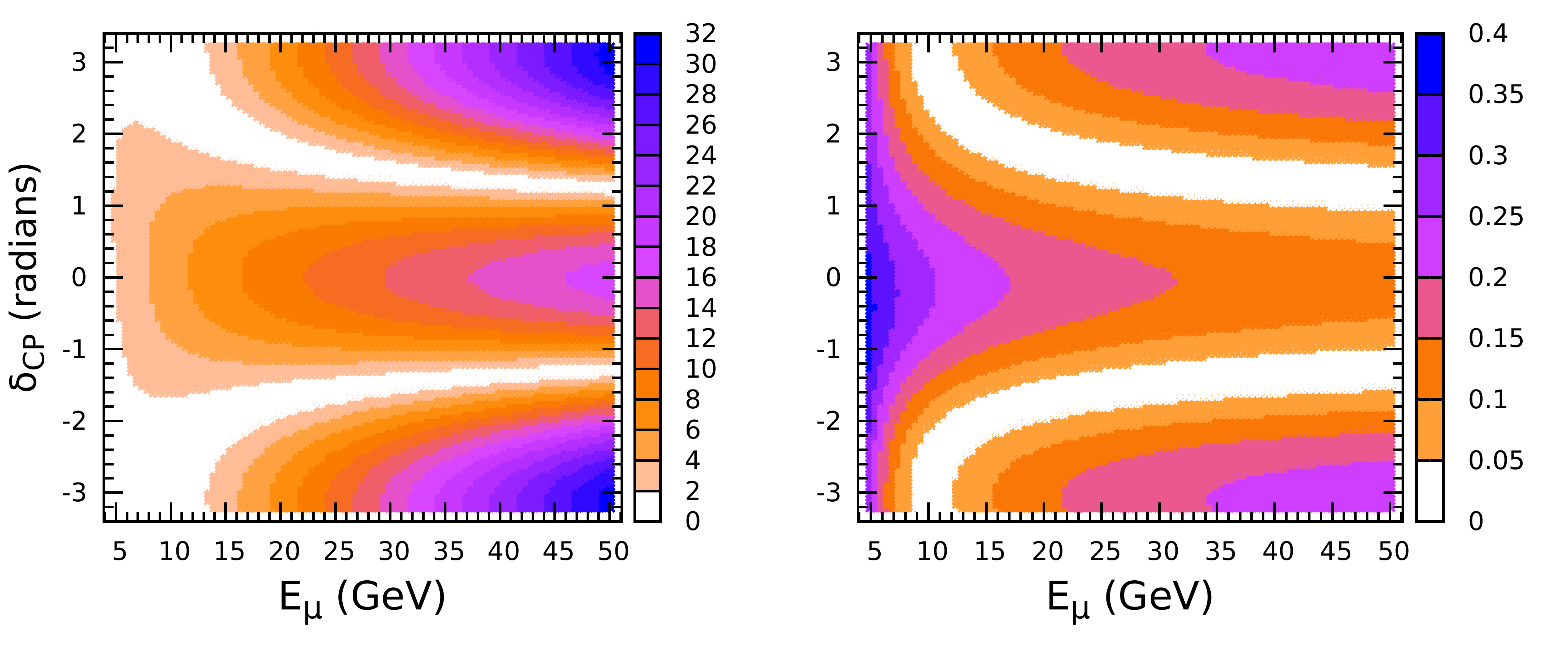}
\label{fig:emu_dcp13}
%\hspace*{\fill}%
\end{figure*}
%%%%%%%%%%%%%%%%%%%%%%%%%%%%%%%%%%%%%%%%%%%%%%%%%%%%%%%%%%%%%%%%%%%%%%%%%%%%%%%%%%%%%%%%%%%%%%%  
%\bigskip
%\hspace*{\fill}%
%%%%%%%%%%%%%%%%%%%%%%%%%%%%%%%%%%% Figure %%%%%%%%%%%%%%%%%%%%%%%%%%%%%%%%%%%%%%%%%%%%%%%%%%%%%%%%%%%%%%%%%%%
\begin{figure*}[h!]
\centering
\caption{(Color online). On LHS is the $N_{e ~\mu}^h$ event rate in $10^3$ units,
RHS shows mass hierarchy asymmetry parameter $N^{asy}$ for the $\nu_e \longrightarrow \nu_\mu$ channel defined in 
Eqn.~(\ref{mhasy}). Here L=812 Km, $\rho = 2.8 ~gm/cm^3$, 
$N_{\mu}^+ = 3 \times 10^{20}$, $N_{KT} = 50 ~KT$, $T= 10 ~Y$. }
%\hspace*{\fill}%
\includegraphics[width=0.97\textwidth]{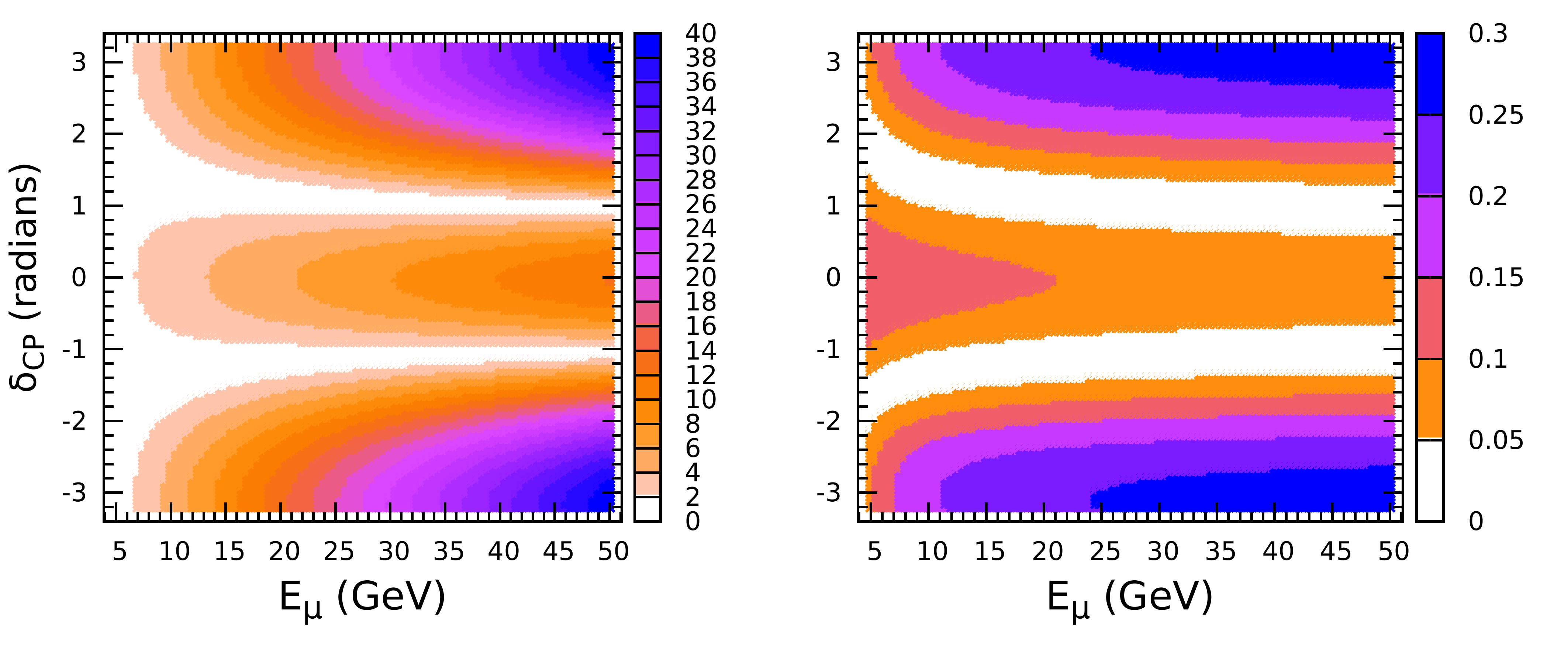}
\label{fig:emu_dcp8}
%\hspace*{\fill}%
\end{figure*}
%%%%%%%%%%%%%%%%%%%%%%%%%%%%%%%%%%%%%%%%%%%%%%%%%%%%%%%%%%%%%%%%%%%%%%%%%%%%%%%%%%%%%%%%%%%%%%%  
 %%%%%
%%
%%%%%%%%%%%%%%%%%%%% Table  %%%%%%%%%%%%%%%%%%%%%%%%%%%%%%%%%%
\setlength{\arrayrulewidth}{0.25 mm}  %% controlls line width 
\setlength{\tabcolsep}{3pt}  %% controlls the width of table or column spacing.
\renewcommand{\arraystretch}{1.5} %% controlls length i.e. row separation.
%%%%%%%%%%%%%%%%%%%%%%%%%%%%%%%%%%%%%%%%%%%%%%
\begin{table*}[htbp!]
%\hspace*{\fill}%
\centering
\caption{This table tabulates Long Base Line (LBL) experimental configurations
         considered in the present work \cite{thzh}, \cite{uno2}.
\label{tab:lblcon}}
\begin{tabular}{ p{3cm} p{2cm} p{2cm} p{4cm} p{3cm}}  \hline
\head{Experiment} &   \head{Baseline} & \head{$\rho_{av}$} & \head{Marginalized range} & $\langle E \rangle \pm \Delta E$\\ 
                  & L (Km)            & ($g/{cm^3}$)       & $\rho$ ($g/{cm^3}$) & (GeV) \cite{thzh}\\
%%%& & & exact & approximate \\ 
\hline
NOVA          & 812  & 2.8  &  2.2 - 3.4 & 2.02 $\pm$ 0.43\\
DUNE (LBNE)   & 1300 & 3.5  &  2.7 - 4.3 & 3.55 $\pm$ 1.38\\ 
UNO-Henderson & 2700 & 3.8  &  3.0 - 4.4 & 6.0 $\pm$ 1.7\\
\hline
\end{tabular}
%%%
\end{table*}
%%%  
  
In Figs.~\ref{fig:emu_dcp27}, \ref{fig:emu_dcp13} and \ref{fig:emu_dcp8}, we have illustrated $\delta_{CP}$ sensitivity
of mass ordering parameter $N_{e ~\mu}^h$ as function of $\mu^+$ beam energy $E_\mu$. In each of the figures, sub-figure 
on LHS represents the mass hierarchy sensitivity parameter $N_{e ~\mu}^h$ defined in Eqn.~(\ref{nabh}) and sub-figure 
on the RHS to the mass hierarchy asymmetry parameter $N^{asy}$ defined in Eqn.~(\ref{mhasy}). As is clear from the 
figures sensitivity is very low for $E_\mu \lessapprox 10$ GeV, while for $E_\mu \gtrapprox 40$ GeV sensitivity becomes 
high. Thus the range $30 \lessapprox E_\mu \lessapprox 50$ GeV is most advantageous range for achiving 
observable sensitivity towards the investigation of $\delta_{CP}$ phase. In case of both UNO (L=2700 Km) and 
DUNE (L= 1300 Km) experiments, the value of mass hierarchy asymmetry $N^{asy}$ parameter is higher for $E_\mu \lessapprox 10$ GeV,
but in case of NO$\nu$A (L= 812 Km) experiment for $E_\mu \gtrapprox 20$ GeV, parameter $N^{asy}$ assumes the higher value. 
From above lines we can conclude to say that experimental setup NO$\nu$A considers the highest precedence over the 
other considered experiments, in respect of having comparable sensitivity and highest $N^{asy}$ (mass ordering difference
to signal ratio) value in the most advantageous energy range $30 \lessapprox E_\mu \lessapprox 50$ GeV.
It is also evident from figures that region can be divided in to two symmetrical halves around $\delta_{CP} =0^0$. Which 
implies that at given muon energy $E_\mu$, both the upper half ($0 \leq \delta_{CP} \leq \pi$) and the 
lower half ($0 \leq \delta_{CP} \leq -\pi$) have equal probability of occurrence. 
This leads to $\delta_{CP}$ upper half and lower half degeneracy. 

We can observe in Figs.~\ref{fig:emu_dcp13} and \ref{fig:emu_dcp8}, that in either of the $\delta_{CP}$ halves (upper or lower), 
there is possibility of existing a given colored region over two different ranges of $\delta_{CP}$ phase. This generates an
internal degeneracy in the given $\delta_{CP}$ half. This in turn hinders the investigation 
of the narrow ranges for the $\delta_{CP}$ phase. But, such type of internal degeneracy is absent for UNO (L= 2700 Km)
experiment, as is evident from Fig.~\ref{fig:emu_dcp27}, where in the given half of $\delta_{CP}$ phase,
a given colored region appears only once, especially in the $30 \lessapprox E_\mu \lessapprox 50$ GeV range.
Which suggests that UNO-Henderson (L=2700 Km) experiment is the most advantageous in order to
investigate narrow ranges of $\delta_{CP}$ phase.

Thus in order to have highest sensitivity towards the $\delta_{CP}$ variations, it is advisable from above 
Figs.~\ref{fig:emu_dcp27},~\ref{fig:emu_dcp13} and \ref{fig:emu_dcp8}, that we should choose 
$E_\mu \simeq 30$ GeV for UNO (L=2700 Km) and $E_\mu \simeq 50$ GeV for DUNE (L=1300 Km) and NO$\nu$A
(L=812 Km) experiments. We also observe that adding anti-neutrino wrong channel 
(i.e. $\overline{\nu}_\mu \rightarrow \overline{\nu}_\mu$) to the neutrino channel ($\nu_e \rightarrow \nu_\mu$)
lowers the $\delta_{CP}$ sensitivity of the experiments. Hence we treat $\overline{\nu}_\mu$ contributions to the 
event rates as the background in the discussion till now. 

%%%%%%%%%%%%%%%%%%%%%%%%%%%%%%%%%%%%%%%%%%%%%%%%%%%%%%%%%%%%%%%%%%%%%%%%%%%%%%%%%%%%%%%%%%%%%%%%%%%%%%%%%%%%%%%%%%%%%%%%%%%%%%%%%%%%%%%%%%%%%%%%
%%%%%%%%%%%%%%%%%%%%%%                                                                                                %%%%%%%%%%%%%%%%%%%%%%%%%%
                      \section{ Octant sensitivity of {{ \fontsize{15}{25}\selectfont \bf $\theta_{23}$ }}}
                                 \label{sec:octsens}
%%%%%%%%%%%%%%%%%%%%%                                                                                                  %%%%%%%%%%%%%%%%%%%%%%%%%
%%%%%%%%%%%%%%%%%%%%%%%%%%%%%%%%%%%%%%%%%%%%%%%%%%%%%%%%%%%%%%%%%%%%%%%%%%%%%%%%%%%%%%%%%%%%%%%%%%%%%%%%%%%%%%%%%%%%%%%%%%%%%%%%%%%%%%%%%%%%%%%%
Chi-square analysis to find the octant of atmospheric angle can be described as following \cite{rg1}, \cite{chi1}, \cite{chi2}, \cite{chi3}
\begin{eqnarray}
 \chi^2 = min_{\xi_k} \left[ \sum_i \frac{\left( n_i^{exp} - \widetilde{n}_i^{~th}\right)^2}{\left(\sigma_i^{stat}\right)^2} 
           + \sum_k^{npull} \xi _k^2 \right]
           \label{chisq}
\end{eqnarray}
where we can choose $\left( \sigma_i^{stat} \right)^2 \approxeq n_i^{exp} $ and
\begin{eqnarray*}
 \widetilde{n}_i^{~th} = n_i^{th} \left[ 1 + \sum_k^{npull} \pi_i^k ~\xi_k \right]
\end{eqnarray*}
where $n_i^{exp}$ is the number of experimental events in the i-th bin for the considered best fit oscillation 
parameters and $n_i^{th}$ is the theoretical number of events in the i-th energy bin for the chosen test oscillation
parameters. Here k runs from 1 to npull, where npull is the number of sources of uncertainty/error. 
The set $\{ \pi_i^k \}$ of parameters is the set of coupling factors, which describe 
the strength of the coupling between the pull $\xi_i^k$ and the observable $n_i^{th}$. The quantities 
$\pi_i^k$ give the fractional rate of change of $n_i^{th}$ due to kth systematic uncertainty.

In our analysis we will include the uncertainties coming from 
\begin{enumerate}
 \item A flux normalization error of 20 \% i.e. $\pi^1 = 0.2$
 \item An overall cross-section uncertainty of 5 \% i.e. $\pi^2 = 0.05$
 \item An overall systematic uncertainty of 5 \% i.e. $\pi^3 = 0.05$
\end{enumerate}
%%%%
The coefficients $\xi_k$'s which minimize the $\chi^2$ function defined in Eqn.~(\ref{chisq}) 
above can be evaluated through the equations
\begin{eqnarray}
  \frac{\partial \chi^2}{\partial \xi_1} = 0 ~; ~~~~~~ \frac{\partial \chi^2}{\partial \xi_2} = 0 ~;  
   ~~~~~~\frac{\partial \chi^2}{\partial \xi_3} = 0 
  \label{partialderi}
\end{eqnarray}
which gives 
\begin{eqnarray}
 \xi_i = \frac{\pi^i b}{1 + a\left( {\pi^1}^2 + {\pi^2}^2 + {\pi^3}^2 \right) }
 ~~~~~~~~ where ~~ i ~= ~1, ~2, ~3
 \label{coefpis}
\end{eqnarray}
with
\begin{eqnarray*}
 a = \sum_i \frac{n_i^{th} ~n_i^{th}}{n_i^{exp}} ~;  ~~~~~ b = \sum_i \left( n_i^{th} - \frac{n_i^{th} ~n_i^{th}}{n_i^{exp}} \right)
\end{eqnarray*}
%%%
On substituting the values from Eqn.~(\ref{coefpis}) in to Eqn.~(\ref{chisq}), 
we have the $\chi^2$ minimized over the pulls, which includes the effects of all systematic and 
theoretical uncertainties as
\begin{eqnarray}
 \chi_{pull}^2 = min_{\xi_k} \left[ \chi^2 \left( \xi_k \right) \right]
 \label{chisqpull}
\end{eqnarray}
As we don't have stringent bounds over the atmospheric mixing angle and mass square difference
(i.e. $\theta_{13}$, $\theta_{23}$, $\Delta m_{23}^2$) and there could be uncertainty in the baseline length 
and hence in the mater density ($\rho$), which we assume to be $\pm$ 5 \%. 
Hence we can marginalized over these parameters to get the final $\chi^2$ as
\begin{eqnarray}
  \chi^2_{marginalized} \equiv \chi^2_{min} = min \Bigg[ \chi^2(\xi_k) \Bigg. 
                       &+& \left( \frac{|\Delta m_{31}^2|^{true} - |\Delta m_{31}^2|^{test}}{\sigma\left( |\Delta m_{31}^2| \right)} \right)^2 
  + \left( \frac{sin^2 2 \theta_{23}^{true} - sin^2 2 \theta_{23}^{test}}{\sigma \left(sin^2 2 \theta_{23}\right)} \right)^2  \nonumber \\
   &+& \left. \left( \frac{sin^2 2 \theta_{13}^{true} - sin^2 2 \theta_{13}^{test}}{\sigma \left(sin^2 2 \theta_{13}\right)} \right)^2 
        + \left( \frac{\rho_0 -\rho}{\sigma(\rho)} \right)^2 \right]
\end{eqnarray}
All other parameters except that of the parameters considered for the marginalization i.e. $\theta_{12}$, $\Delta m_{13}^2$ 
and phase $\delta_{CP}$ are kept at the best fit values in $n_i^{th}$ calculations, while $n_i^{exp}$ has been calculated
for the best fit/true oscillation parameters. Thus
\begin{eqnarray}
 n_i^{exp} &\equiv& n_i^{exp}\left( {\theta_{12}}^{true}, ~{\theta_{23}}^{true}, ~{\theta_{13}}^{true}, ~{\Delta m _{12}^2}^{true}, ~
 {\Delta m _{13}^2}^{true}, ~{\delta_{CP}}^{true}, ~\rho^{true} \right)   \nonumber  \\
 %%%%%%
 n_i^{th} &\equiv& n_i^{th}\left( {\theta_{12}}^{true}, ~{\theta_{23}}^{test}, ~{\theta_{13}}^{test}, ~{\Delta m _{12}^2}^{true}, ~
 {\Delta m _{13}^2}^{test}, ~{\delta_{CP}}^{true}, ~\rho^{test} \right)
 \label{paraent}
\end{eqnarray}
%%
%%%
%%%%%%%%%%%%%%%%%%%% Table  %%%%%%%%%%%%%%%%%%%%%%%%%%%%%%%%%%
\setlength{\arrayrulewidth}{0.25 mm}  %% controlls line width 
\setlength{\tabcolsep}{25pt}  %% controlls the width of table or column spacing.
\renewcommand{\arraystretch}{1.7} %% controlls length i.e. row separation.
%%%%%%%%%%%%%%%%%%%%%%%%%%%%%%%%%%%%%%%%%%%%%%
\begin{table*}[htbp!]
\centering
\caption{
         Chosen benchmark values of the marginalized oscillation parameters and 
their 1$\sigma$ estimated errors. The last row represents the chosen values 
for the Earth matter density at given baseline length. 
\label{tab:bench1sig}  
        }
\begin{tabular}{ p{5cm} p{4cm}}  \hline
\head{Marginalized Parameter} &   \head{1 $\sigma$ error}  \\  
%%%& & & exact & approximate \\ 
\hline
$|\Delta m_{31}^2|_{NH}^{true} = 2.48 \times 10^{-3} eV^2$  & $\sigma\left( |\Delta m_{31}^2| \right) = 15 \% $ \\
$|\Delta m_{31}^2|_{IH}^{true} = 2.44 \times 10^{-3} eV^2$  & $\sigma\left( |\Delta m_{31}^2| \right) = 15 \% $ \\
$\left[sin^2 2 \theta_{23}\right]_{NH}^{true} = 0.99$ & $\sigma \left(sin^2 2 \theta_{23}\right) = 1 \%$ \\
$\left[sin^2 2 \theta_{23}\right]_{IH}^{true} = 0.98$ & $\sigma \left(sin^2 2 \theta_{23}\right) = 1 \%$ \\ 
$\left[sin^2 2 \theta_{13}\right]_{NH}^{true} = 0.091$ & $\sigma \left(sin^2 2 \theta_{13}\right) = 10 \%$  \\
$\left[sin^2 2 \theta_{13}\right]_{IH}^{true} = 0.093$ & $\sigma \left(sin^2 2 \theta_{13}\right) = 10 \%$  \\
$\rho_0$ & $\sigma(\rho) = 5 \%$  \\
\hline
\end{tabular}
%%%
\end{table*}
%%%%%%%%%%%%%%%%%%%%%%%%%%%%%%%%%%%5
\begin{figure*}[h!]
\centering
\caption{(Color online). Chi square fit for the chosen experiments,
$N_{\mu}^+ = 3 \times 10^{20}$, $N_{KT} = 10 ~KT$, $T= 1 ~Y$ for the LHS plot 
and $N_{KT} = 50 ~KT$, $T= 10 ~Y$ for the RHS plot. $\chi^2$ has weak dependence on
$\delta_{CP}$ phase, hence we choose $\delta_{CP} = 0$. $E_\mu = 30$ GeV for UNO (L=2700 Km)
and $E_\mu = 50$ GeV for DUNE (L=1300 Km) \& NO$\nu$A (L=812 Km).}
\includegraphics[width=0.95\textwidth]{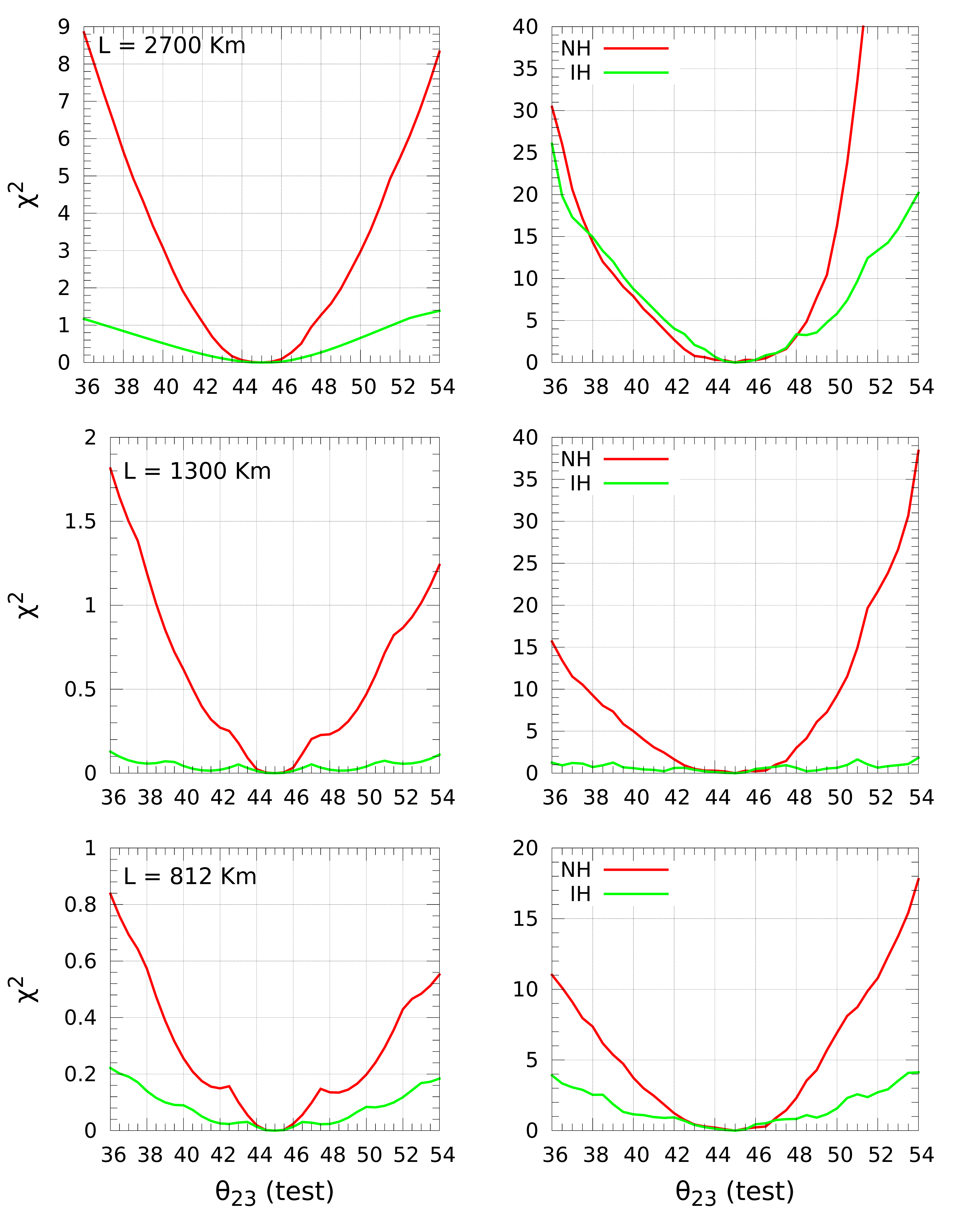}
\label{fig:chisq23}
\end{figure*}   
%%%
In Fig.~\ref{fig:chisq23}, we have illustrated $\chi^2$ analysis for the three considered experiments 
to investigate the octant sensitivity of atmospheric mixing angle $\theta_{23}$. In order to 
find the minimum possible value of $\chi^2$ over the $3~\sigma$ range of mixing parameters and 
10~\% possible density fluctuations, we have marginalized $\chi^2$ deviations with respect the test
parameters $\theta_{23}$, $\theta_{13}$, $\Delta m_{13}^2$ and $\rho$. It is observable from the figure 
that for detector configuration 10 KTY (i.e. sub-figures on LHS column), octant sensitivity is very low in 
comparison to the 500 KTY (i.e. sub-figures on RHS column) detector configuration. Octant sensitivity
in case of NH is much better than the IH case, except that UNO experiment where for the 500 KTY detector 
configuration for both NH and IH case octant sensitivity is almost same for the lower octant. In the latter 
case octant sensitivity up to 2$\sigma$ (i.e. $\chi^2 = 4$) CL level is same for both the hierarchies.
%%%%%
\setlength{\arrayrulewidth}{0.35 mm}  %% controlls line width 
\setlength{\tabcolsep}{2pt}  %% controlls the width of table or column spacing.
\renewcommand{\arraystretch}{2.1} %% controlls length i.e. row separation.
%%%%%%%%%%%%%%%%%%%%%%%%%%%%%%%%%%%%%%%%%%%%%%
\begin{table}[h!]
  \centering
  \caption{Ranges of $\theta_{23}$ corresponding to the 1$\sigma$ ($\chi^2 = 1$), 2$\sigma$ ($\chi^2 = 4$) and 3$\sigma$ ($\chi^2 = 9$)
           CL's., from figure~\ref{fig:chisq23}. The detector configuration 10 KTY corresponds to $N_{KT} = 10$ KT and time period T=1 Year,
           while 500 KTY detector configuration corresponds to $N_{KT} = 50$ KT and time period T=10 Years. The dash symbol 
           indicates that the value of $\chi^2$ is less than the respective value of confidence level (CL) over the 
           full $36^0 \leq \theta_{23} \leq 54^0$ range.}
  \begin{tabular}{|c|c|c|c||c|c|c||c|c|c||c|c|c|}
    \hline
    \multirow{4}{*}{\textbf{Experiment}} & \multicolumn{6}{c||}{\textbf{Detector Configuration (10 KTY)}} 
                                                                     & \multicolumn{6}{c|}{\textbf{Detector Configuration (500 KTY)}}  \\
     \cline{2-13}
    & \multicolumn{6}{c||}{\textbf{{\bf \fontsize{15}{31}\selectfont $\theta_{23}$} range for}} 
    & \multicolumn{6}{c|}{\textbf{{\bf \fontsize{15}{31}\selectfont$\theta_{23}$} range for}}\\
    \cline{2-13}
    & \multicolumn{3}{c||}{\textbf{{\bf \fontsize{15}{31}\selectfont $\chi^2$}(IH)}} 
    &  \multicolumn{3}{c||}{\textbf{{\bf \fontsize{15}{31}\selectfont $\chi^2$}(NH)}} 
                                                 & \multicolumn{3}{c||}{\textbf{{\bf \fontsize{15}{31}\selectfont $\chi^2$}(IH)}} 
                                                 &  \multicolumn{3}{c|}{\textbf{{\bf \fontsize{15}{31}\selectfont $\chi^2$}(NH)}}\\
         \cline{2-7} \cline{8-13}
        & \textbf{{\bf \fontsize{16}{31}\selectfont $1 ~\sigma$}} & \textbf{{\bf \fontsize{16}{31}\selectfont $2 ~\sigma$}} 
        & \textbf{{\bf \fontsize{16}{31}\selectfont $3 ~\sigma$}} 
           & \textbf{{\bf \fontsize{16}{31}\selectfont $1 ~\sigma$}} & \textbf{{\bf \fontsize{16}{31}\selectfont $2 ~\sigma$}} 
           & \textbf{{\bf \fontsize{16}{31}\selectfont $3 ~\sigma$}} 
        & \textbf{{\bf \fontsize{16}{31}\selectfont $1 ~\sigma$}} & \textbf{{\bf \fontsize{16}{31}\selectfont $2 ~\sigma$}} 
        & \textbf{{\bf \fontsize{16}{31}\selectfont $3 ~\sigma$}} 
           & \textbf{{\bf \fontsize{16}{31}\selectfont $1 ~\sigma$}} & \textbf{{\bf \fontsize{16}{31}\selectfont $2 ~\sigma$}} 
           & \textbf{{\bf \fontsize{16}{31}\selectfont $3 ~\sigma$}}  \\
    \hline  \hline
\textbf{NO$\nu$A (812 Km)}  & --- & --- & --- &   36--54 & --- & --- &      41--48 & 36--54 & ---   & 43--47 & 40--49 &  37--51  \\ 
    \hline
\textbf{DUNE (1300 Km)}     & --- & --- & --- &   39--53 & --- & --- &      36--54 & --- & ---   & 42--47 & 40--48 &  38--50  \\  
    \hline
\textbf{UNO (2700 Km)}      & 37--51 & --- & --- &   42--47 & 39--51 & 36--54 &      44--47 & 42--49 & 40--51   & 43--47 & 41--48 &  39--49  \\  
    \hline
  \end{tabular}
  \label{tab:chisq}
\end{table}

We can observe from Table~\ref{tab:chisq}, for detector configuration 10 KTY, octant sensitivity in case of 
inverted hierarchy (IH) is almost negligible in comparison to normal hierarchy (NH) case.
Only for the UNO (L=2700 Km) experiment, 1$\sigma$ (IH) sensitivity can be possible, 
which is very less in comparison to the corresponding NH case. We also observe that in the NH case, as the base line 
length increases, $\theta_{23}$-octant sensitivity also increases. Experiment UNO provides the highest 
sensitivity, which is three times of the DUNE and four times of the NO$\nu$A sensitivity at 1$\sigma$ CL,
as is clear from 5th column of Table~\ref{tab:chisq}. Hence for detector configuration 10 KTY, only 
UNO experiment provides the opportunity to investigate $\theta_{23}$ octant up to 1$\sigma$ CL. 
More explicitly, we can say that with this detector configuration only in the case of NH, the 
investigation of $\theta_{23}$ octant is possible. But if nature has chosen IH for the 
neutrino mass spectra, then octant determination is not possible in this case, 
as sensitivity is too low to be observed experimentally.

In the case of detector configuration of 500 KTY, there is possibility to detect octant in the IH case too, but still 
sensitivity in the NH case dominates the sensitivity in the IH case. If we compare 1$\sigma$ sensitivity
in the IH case, UNO sensitivity is twice that of NO$\nu$A sensitivity, as is clear from 8th column.
But DUNE experiment exhibits negligible octant sensitivity in the IH case.
Also for UNO experiment in case of both NH and IH case, $\theta_{23}$ octant sensitivity
is almost same up to 2$\sigma$ CL. For the DUNE experiment, it is also observable that  
in the lower octant ($36 \leqslant \theta_{23} \leqslant 45$) up to 5$\sigma$ CL for both NH and IH
case octant sensitivity is almost same. As is clear from columns 11, 12 and 13, the respectively
1$\sigma$, 2$\sigma$ and 3$\sigma$ level sensitivities are almost equal for all the three experiments.
Now if we look 
on the RHS column of Fig.~\ref{fig:chisq23}, for UNO experiment sensitivities up to 5$\sigma$ CL 
can be achieved. Also for both UNO and DUNE experiments in the 
upper octant ($45 \leqslant \theta_{23} \leqslant 54$), the NH sensitivities $\geqslant 5 \sigma$ CL can be achieved.
But for NO$\nu$A experiment we can achieve sensitivity up to 3$\sigma$ CL in this regard. Thus we can conclude 
to say that all the three experiments provide almost equal sensitivities to investigate the $\theta_{23}$ octant 
up to 3$\sigma$ CL. Experiment UNO, provides equal opportunities to investigate $\theta_{23}$ octant 
in the both NH and IH case up to 2$\sigma$ CL and up to 5$\sigma$ CL in the lower octant.

Now if we compare two detector configurations, then in the case of IH for the UNO experiment, the 
octant sensitivity of 500 KTY detector configuration is approximately 5 times that of 10 KTY 
configuration up to 1$\sigma$ CL, as is evident from the last row of 2nd and 8th columns. For the remaining
experiments and higher order CL for the 10 KTY detector configuration in the IH case, we don't expect 
any sensitivity. In the NH case, experiment NO$\nu$A (L=812 Km) exhibits negligible sensitivity
for 10 KTY detector configuration in comparison to the 500 KTY detector configuration. 
For DUNE (L=1300 Km) experiment, detector configuration 500 KTY has a sensitivity which is 3 times of the 
corresponding 10 KTY configuration sensitivity up to 1$\sigma$ CL, higher order CL sensitivities can be achieved
in the former case, while for 10 KTY configuration these sensitivities are negligible. In the 
UNO (L=2700 Km) experiment both detector configurations exhibit almost equal sensitivity 
up to 1$\sigma$ CL, but the sensitivity of 500 KTY configuration is twice of the 10 KTY configuration 
at 2$\sigma$ and 3$\sigma$ CL's of the $\chi^2$ parameter.
%%%%%
\begin{figure*}[htbp!]
\centering
\caption{(Color online). Total chi square for the synergistically combined NO$\nu$A+DUNE+UNO data with priors. 
Marginalized over $7.4^0 \leq \theta_{13} \leq 10^0$. We consider on axis neutrino beam.}
\includegraphics[width=0.75\textwidth]{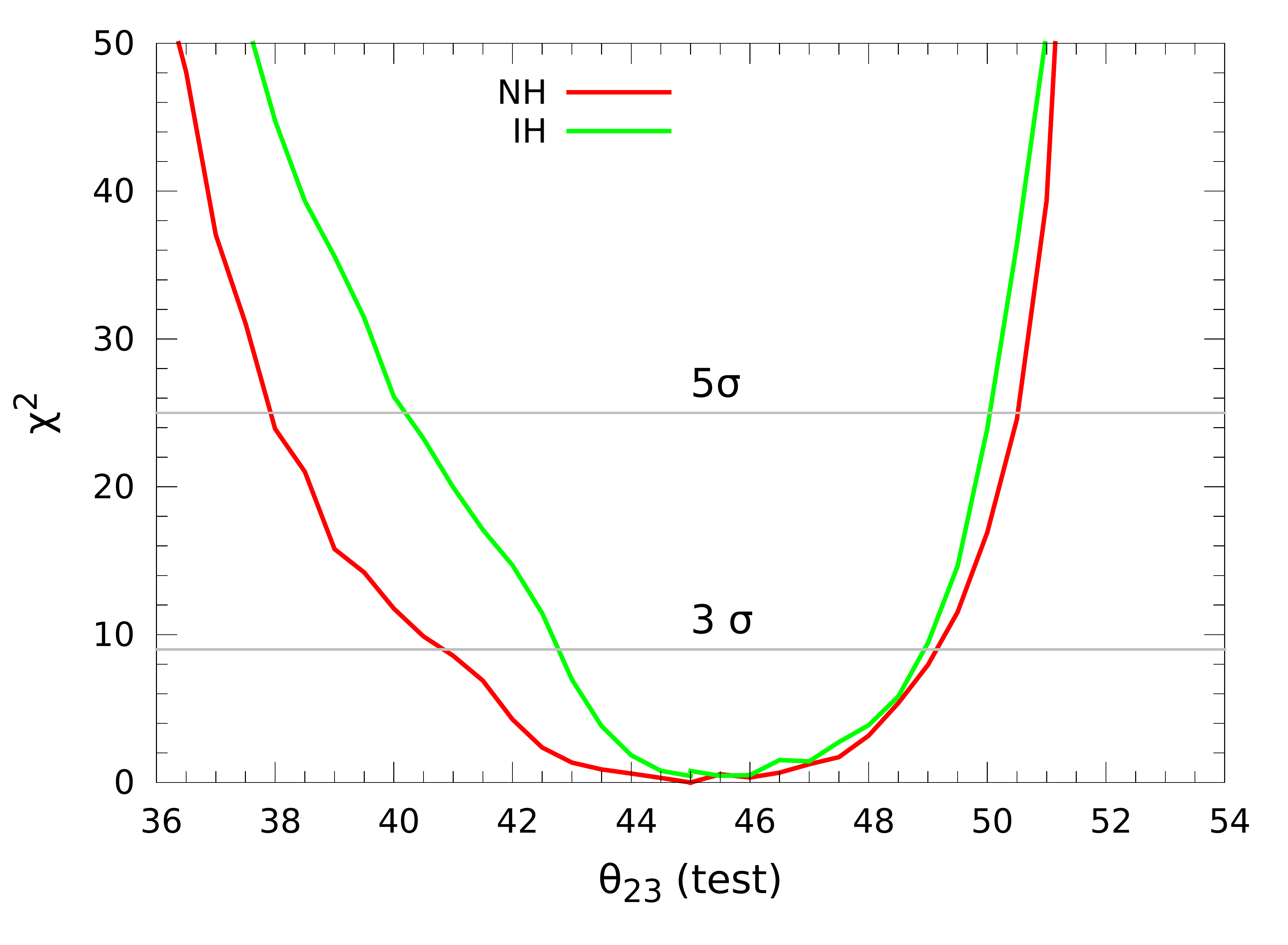}
\label{fig:totchi}
\end{figure*}   
 %%%%

In Fig.~\ref{fig:totchi}, a combined analysis of three considered experiments (NO$\nu$A+DUNE+UNO) has been 
illustrated. In this case, we choose 
\begin{eqnarray*}
 \chi^2_{prior} = \left(\frac{sin^2 2 \theta_{13}^{true} - sin^2 2 \theta_{13}^{test}}{\sigma \left(sin^2 2 \theta_{13}\right)} \right)^2        
\end{eqnarray*}
%%%
It is evident from Fig.~\ref{fig:totchi}, synergistic combination of NO$\nu$A+DUNE+UNO data provides the opportunity
to investigate octant up to 5$\sigma$ or even higher C.L. We notice that, in the upper test octant for both NH and IH cases, octant 
sensitivity is almost same, while in the lower test octant, the octant sensitivity for IH case is almost
twice that for the NH case. A hint towards this behavior comes from RHS sub-figure of first row of Fig.~\ref{fig:chisq23},
where IH octant sensitivity is bit higher that of NH sensitivity up to 3$\sigma$ level.
 While the respective octant sensitivity in the IH case 
for other two experimental configurations is negligible in comparison to the respective IH sensitivity 
for the UNO experiment. This can be attributed to the fact that,
at the UNO base line the amplitude of transition probability for the 
NH and IH are almost comparable to each other, that in turn make the octant sensitivity also 
comparable for both NH and IH cases in the upper octant.
Hence it is UNO octant sensitivity that dominates in the combined data of three 
experiments. The $\chi^2$ fit for the combined data represented in the Fig.~\ref{fig:totchi} have been 
marginalized over the $\theta_{13}$ mixing angle only. The marginalization over the other parameters
may improve the results further, but not to much large extent. Although UNO, 500 KTY experimental configuration 
provides observable octant sensitivity for both type of hierarchies, the synergistic addition of 
three experiments data further enhances the sensitivity over both the test octants appreciably.

\section{Analysis of contour plots and precision in {{ \fontsize{15}{25}\selectfont \bf $\theta_{23}$}}
           and {{ \fontsize{15}{25}\selectfont \bf $\delta_{CP}$}}} 
%%%%
%%%%%%%
\begin{figure*}[htbp!]
\centering
\caption{(Color online). Contour plot for the chi square in the $\theta_{23}$ and $\delta_{cp}$ plane for NO$\nu$A(L=812 Km) experiment with priors. 
Marginalized over $7.4^0 \leq \theta_{13} \leq 10^0$, $2 \times 10^{-3} ~eV^2\leq \Delta m_{13}^2 \leq 3 \times 10^{-3} ~ eV^2$) and 
$2.2 \leq \rho_{avg} \leq 3.4 ~ gm/{cm}^3$. The successive rows correspond to $\theta_{23} = 38^0, ~41^0, ~49^0, ~52^0$ and successive columns to
$\delta_{CP}= 90^0, ~0^0, ~-90^0$ respectively.}
\includegraphics[width=0.99\textwidth]{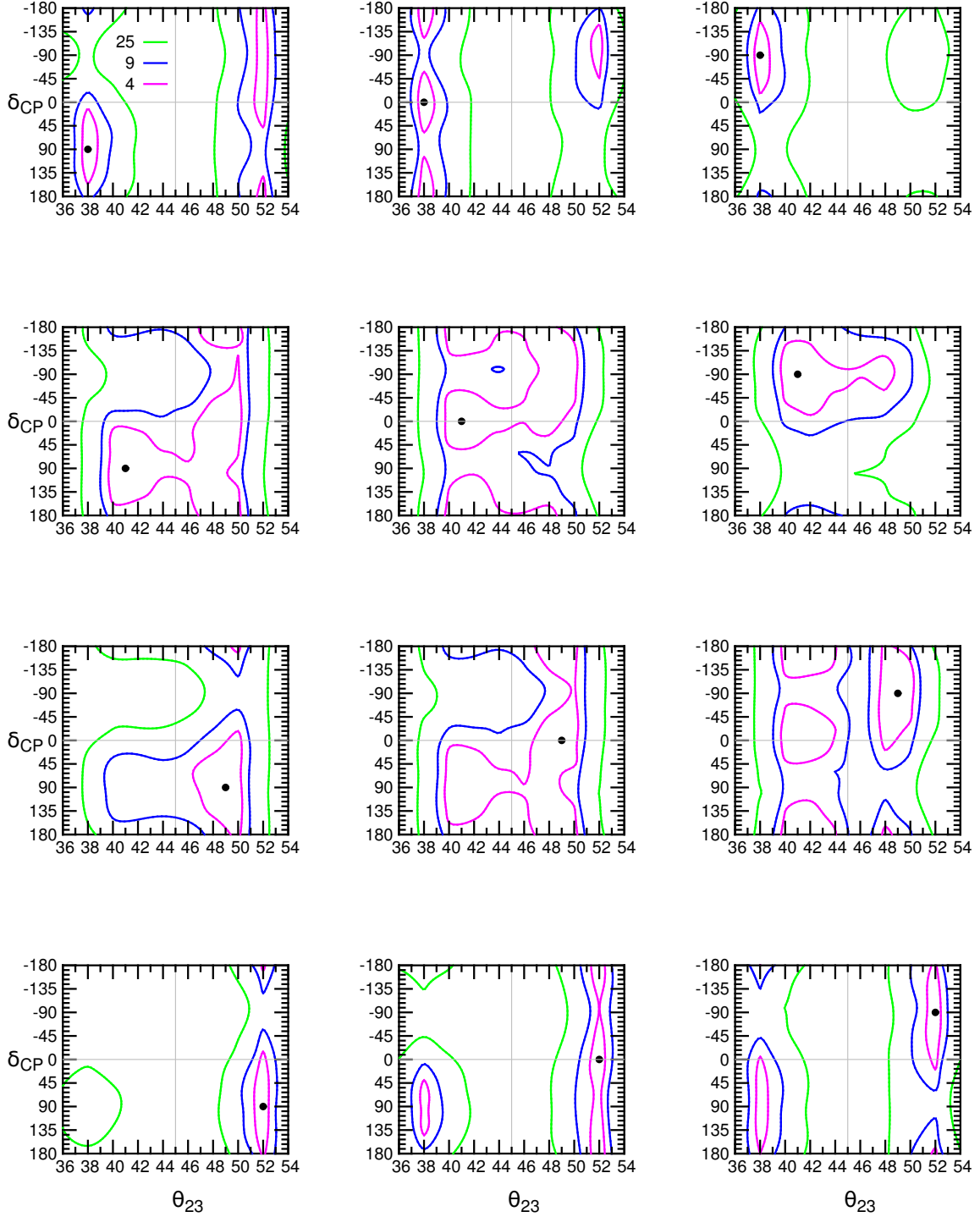}
\label{fig:chicp8}
\end{figure*}   
 %%%% 
%%%
\begin{figure*}[htbp!]
\centering
\caption{(Color online). Contour plot for the chi square in the $\theta_{23}$ and $\delta_{cp}$ plane for DUNE(L=1300 Km) experiment with priors. 
Marginalized over $7.4^0 \leq \theta_{13} \leq 10^0$, $2 \times 10^{-3} ~eV^2\leq \Delta m_{13}^2 \leq 3 \times 10^{-3} ~ eV^2$) and 
$2.7 \leq \rho_{avg} \leq 4.3 ~ gm/{cm}^3$. The rest is same as in Fig.~\ref{fig:chicp8} above.}
\includegraphics[width=0.99\textwidth]{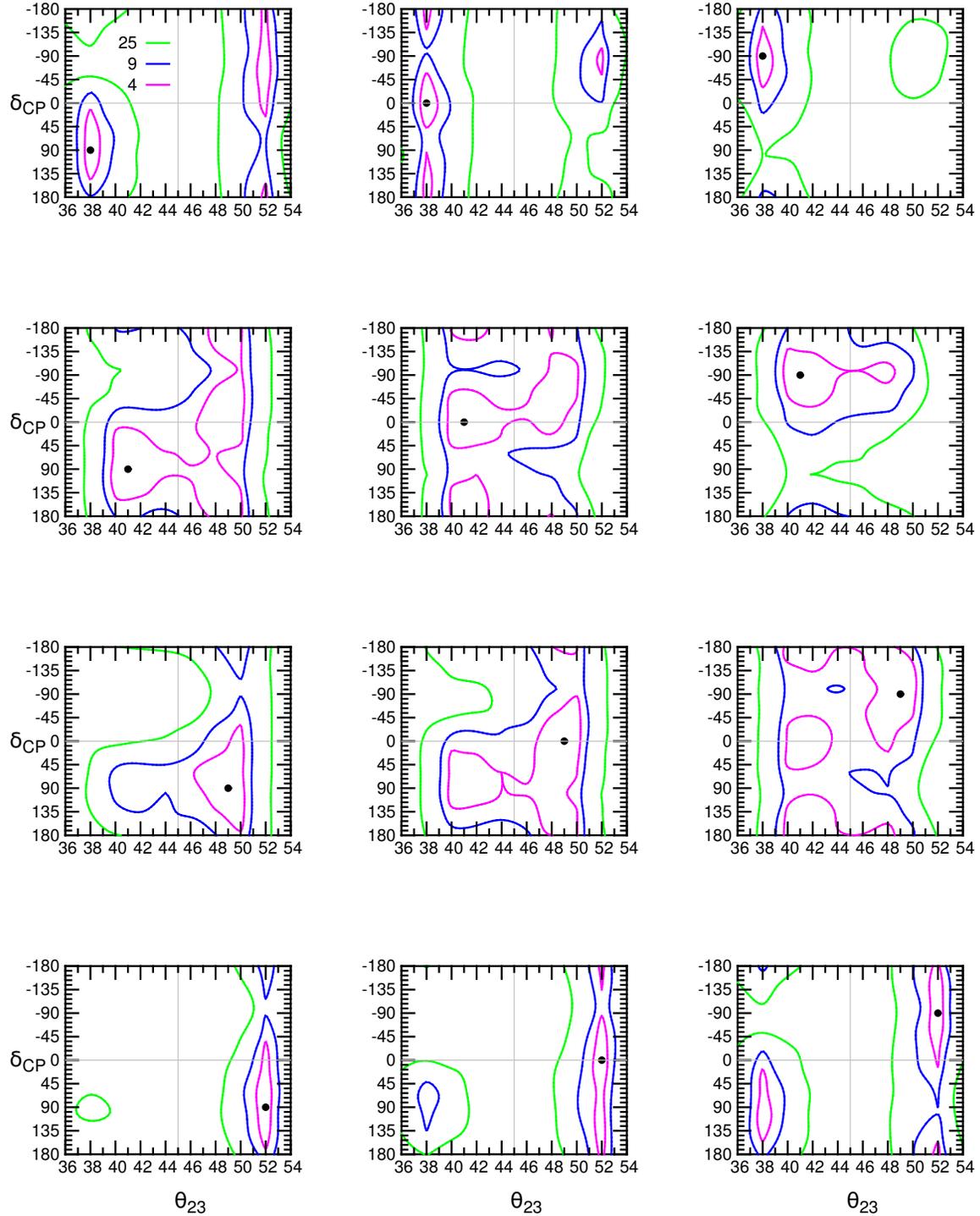}
\label{fig:chicp13}
\end{figure*}   
 %%%% 
%%%% 
 \begin{figure*}[htbp!]
\centering
\caption{(Color online). Contour plot for the chi square in the $\theta_{23}$ and $\delta_{cp}$ plane for UNO(L=2700 Km) experiment with priors. 
Marginalized over $7.4^0 \leq \theta_{13} \leq 10^0$, $2 \times 10^{-3} ~eV^2\leq \Delta m_{13}^2 \leq 3 \times 10^{-3} ~ eV^2$) and 
$3.0 \leq \rho_{avg} \leq 4.4 ~ gm/{cm}^3$. The rest is same as in Fig.~\ref{fig:chicp8} above.}
\includegraphics[width=0.99\textwidth]{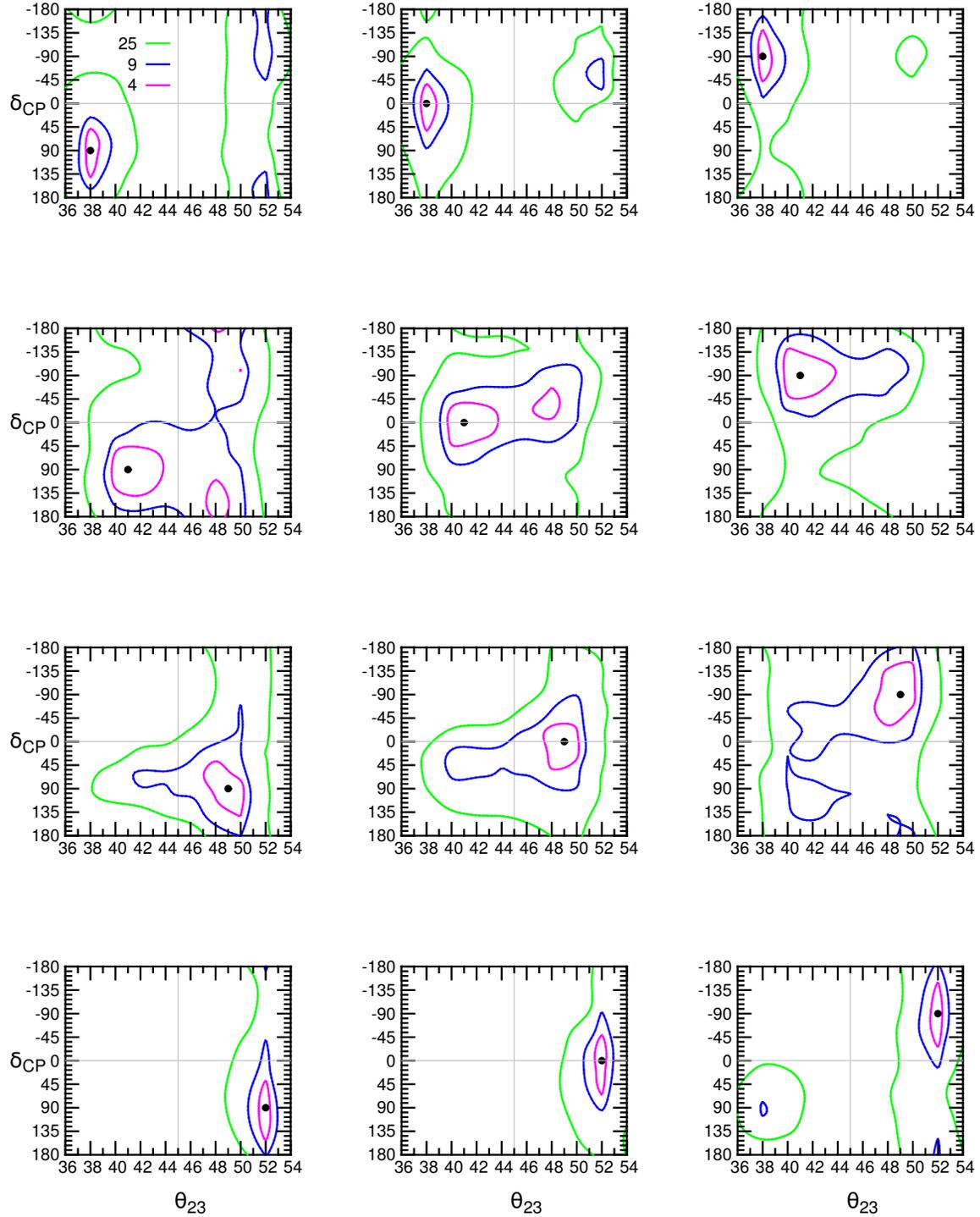}
\label{fig:chicp27}
\end{figure*}   
 %%%% 
%%%
%%%
\begin{figure*}[htbp!]
\centering
\caption{(Color online). Total of chi square for experiments (NO$\nu$A+DUNE+UNO) with priors. 
Marginalized over $7.4^0 \leq \theta_{13} \leq 10^0$ and $2 \times 10^{-3} ~eV^2\leq \Delta m_{13}^2 \leq 3 \times 10^{-3} ~ eV^2$).
The successive rows correspond to $\theta_{23} = 38^0, ~41^0, ~49^0, ~52^0$ and successive columns to
$\delta_{CP}= 90^0, ~0^0, ~-90^0$ respectively. The magenta, blue and green colored contours respectively correspond to
2$\sigma$, 3$\sigma$ and 5$\sigma$ C.L. of $\chi^2$ fit.}
\includegraphics[width=0.99\textwidth]{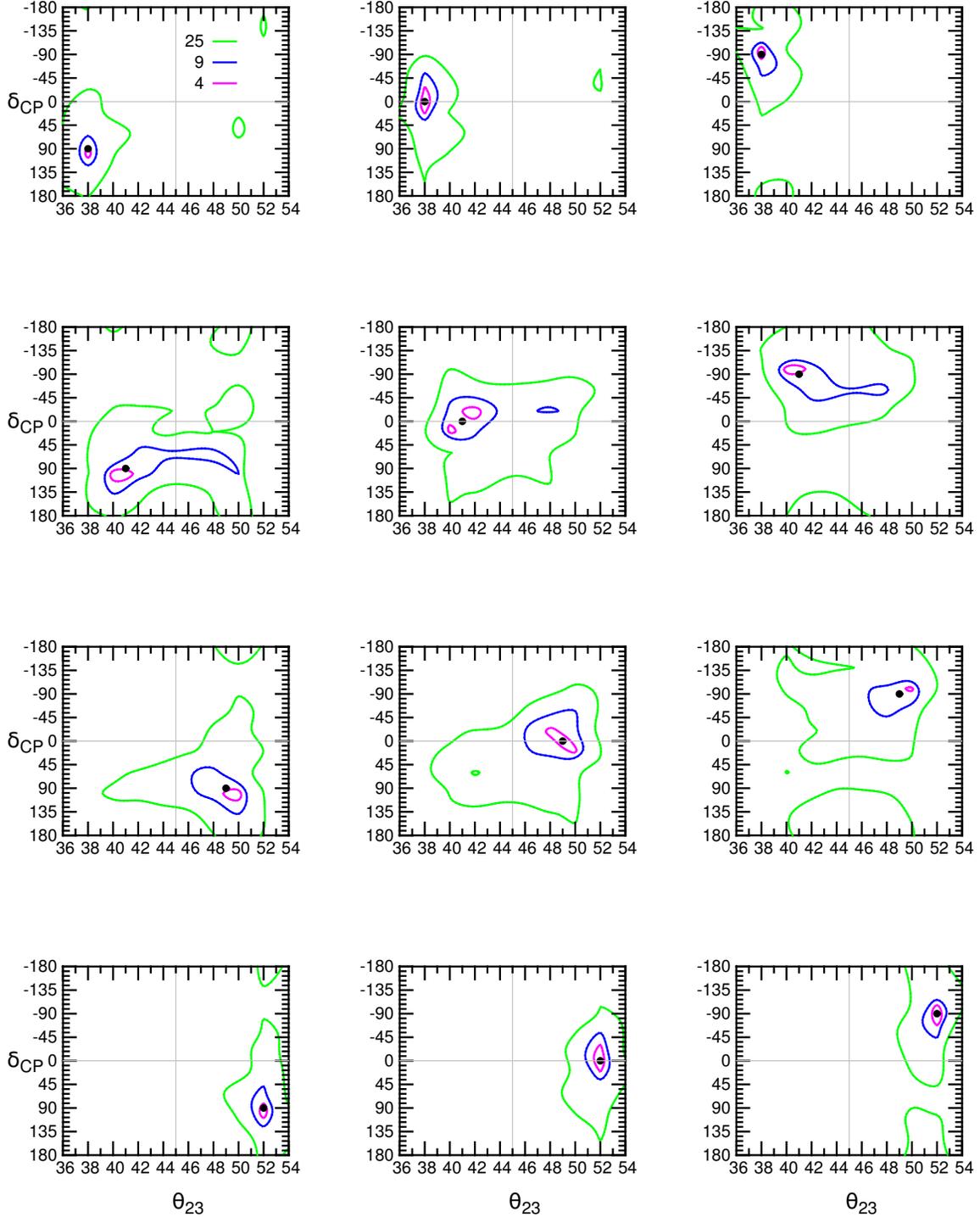}
\label{fig:totchicp}
\end{figure*}   
 %%%% 
 %%%%
In Figs.~\ref{fig:chicp8}, \ref{fig:chicp13} and \ref{fig:chicp27}, the contour plots in the $\theta_{23} - \delta_{CP}$ plane
has been drawn for the three experimental configurations viz. NO$\nu$A(L=812 Km); DUNE(L=1300 Km) and UNO(L=2700 Km) respectively.
In Fig.~\ref{fig:totchicp}, similar type of plots for the combined data of all the three considered experiments 
(i.e. NO$\nu$A + DUNE + UNO) have been depicted. The true $\delta_{CP}$ values chosen are $\pm ~90^0$
corresponding to maximum CP violation and $0^0$ corresponding to
CP conservation. In our analysis we will use W for wrong; R for right; H for hierarchy; O for octant and word
{\it combined data} for the synergistic addition of data from three experiments. 
In these figures the following generic features can be noted: 
\begin{enumerate}
 \item At $\delta_{CP}=90^0$ (i.e. first column of figures), we note that both WO-R$\delta_{CP}$ and WO-W$\delta_{CP}$ 
 solutions along with true solution are present. 
 \begin{enumerate}
  \item  At $\theta_{23}= 38^0$ for both NO$\nu$A and DUNE experiments, LO(lower octant $\theta_{23} \leq 45^0$) and 
 UO(upper octant $\theta_{23} \geq 45^0$) solutions are distinct and right quadrants (RQ) i.e. upper right quadrant (URQ)
 and lower right quadrant (LRQ) solutions (especially 3$\sigma$, 5$\sigma$)
 merge to each other that results in the continuous solution covering entire $\delta_{CP}$ range. While 2$\sigma$
 solution is discrete. Correspondingly in case of UNO experiment the 2$\sigma$ and 3$\sigma$ solutions are distinct, 
 only 5$\sigma$ solution in the RQ merges to give continuous solution. Now if we compare with the results from 
 combined data shown in Fig.~\ref{fig:totchicp}, all the discrete solutions get removed up to 3$\sigma$ level and
 5$\sigma$ solutions become distinct with reduced sizes. Very small size of contour regions 
 provides high precision in the $\theta_{23}$ and $\delta_{CP}$ measurements.
 
 \item At $\theta_{23}= 41^0$ in case of both NO$\nu$A and DUNE experiments different discrete solution regions get 
 merge with true solutions giving continuous solution even at 2$\sigma$ level. In this case almost whole $\delta_{CP}$
 range get covered. This type of discussion is also true at $\theta_{23} = 49^0$ except at 2$\sigma$ C.L., where 
 only about half range of $\delta_{CP}$ is covered. The convergence of different discrete solutions to continuous 
 solution can be attributed to the fact that both $\theta_{23}=41^0$ and $49^0$ lie in the proximity to maximal mixing.
 For more details see \cite{atmex17} and references therein. The synergistic addition of data for 
 all the three experiments depicted in Fig.~\ref{fig:totchicp} 
 reveals that up to 2$\sigma$ level all the discrete and continuous regions 
 get resolved to narrow regions. But at and above 3$\sigma$ level for $\theta_{23}= 41^0$ continuous solutions
 extends to the WO and W$\delta_{CP}$ quadrant regions, while for $\theta_{CP}=49^0$ up to 3$\sigma$ level all the 
 discrete solutions get resolved, as is evident from the very small region enclosed by the blue colored 
 contour in the LRQ in Fig.~\ref{fig:totchicp}.
 
 \item At $\theta_{23}= 52^0$ for all the three experiments up to 3$\sigma$ level no WO-R$\delta_{CP}$ solutions appear though there
 is WO-W$\delta_{CP}$ solutions in case of both NO$\nu$A and DUNE experiments but not in the UNO experiment. At 5$\sigma$ 
 level whole of $\delta_{CP}$ range is allowed and there appears WO solutions in case of first two experiments but 
 in the UNO experiment such solutions are absent. It is evident from the analysis of combined data in Fig.~\ref{fig:totchicp},
 up to 3$\sigma$ C.L. all the degenerate solutions get resolved. Very small size of contour regions 
 facilitates high precision in the $\theta_{23}$ and $\delta_{CP}$ measurements.
 \end{enumerate}
 
 \item At $\delta_{CP}= -90^0$ (i.e. third column of figures), a discussion similar to the $\delta_{CP}= 90^0$ holds good, except that
       in this case RO-W$\delta_{CP}$ solutions also appear with true degenerate solutions.
 
 \item At $\delta_{CP}= 0^0$ (i.e. second column of figures), in case of NO$\nu$A and DUNE experiments 
       for $\theta_{23}=38^0$ and $52^0$ solutions are discrete but for $\theta_{23}= 41^0$ and $49^0$ 
       WO solutions merge with true solution. In case of UNO experiment i.e. Fig.~\ref{fig:chicp27}
       for $\theta_{23}= 38^0$ and $52^0$ multiple solutions are discrete up to 5$\sigma$ level. At
       $\theta_{23}= 41^0$ and $49^0$ solutions are discrete up to 2$\sigma$ level, but at and above 
       3$\sigma$ level all the solutions merge with true solutions to give continuous solution. Also 
       for $\theta_{23}= 52^0$, there are no WO solutions i.e. these have been resolved due to the inclusion of
       large matter effects at longer UNO base line. Now if we compare our results with combined data results in 
       Fig.~\ref{fig:totchicp}, all the discrete and continuous solutions for different degenerate  solutions (i.e. different 
       $\theta_{23}$ and $\delta_{CP}$ true points) get resolved especially up to 3$\sigma$ level. Also the combined data improves the 
       precision of $\theta_{23}$ and $\delta_{CP}$, as is evident from the comparatively small regions enclosed 
       by contour curves.
       
\end{enumerate}

Apart from the above features, the following important points can be observed from the
figures:
\begin{enumerate}
 \item The synergistic addition of three experimental data (i.e. Fig.~\ref{fig:totchicp})
 helps in reducing wrong-octant and W$\delta_{CP}$ extensions even at 5$\sigma$ level. The WO-R$\delta_{CP}$ and WO-W$\delta_{CP}$ solutions 
 also get significantly reduced in size by synergistic addition of experimental data, as can be seen from 
 first row LHS sub-figures. This is due to the fact that for different
 experiments these solutions occur at different $\delta_{CP}$ values. 
 
 \item The synergistic addition of the data set also significantly reduces the size of solutions appearing at 
 5$\sigma$ level especially for $\theta_{23}= 38^0$ and $52^0$, but at $\theta_{23}= 41^0$ and $49^0$ solutions 
 still have comparatively large size at this C.L.
\end{enumerate}

It is noteworthy that the allowed area in the test $\theta_{23} - \delta_{CP}$ plane also gives an idea about
the precision of these two parameters. In general the presence of multiple degenerate solutions leads 
to a worse precision (a larger width of the allowed area) in these parameters. The synergistic 
addition of data for three experiments not only removes the discrete solutions but also reduces the 
size of contour regions around the true point, which in turn provides high precision of these two parameters.
The precision in these parameters can be quantified using the following formulas:
\begin{eqnarray}
 \delta_{CP}^{precision} &=& \frac{\delta_{CP}^{max} - \delta_{CP}^{min}}{360^0} \times 100 ~\%     \\
 \theta_{23}^{precision} &=& \frac{\theta_{23}^{max} - \theta_{23}^{min}}{\theta_{23}^{max} + \theta_{23}^{min}} \times 100 ~\%
 \label{prcision}
\end{eqnarray}
%%%%%%%%%%%%%%%%%%%%%%%%%%%%%%
\setlength{\arrayrulewidth}{0.35 mm}   
\setlength{\tabcolsep}{8pt}  
\renewcommand{\arraystretch}{1.5} 
%%%%%%%%%%%%%%%%%%%%%%%%%%%%%%%%%%%%%%%%%%%%%%
\begin{table}[!htb]
  \centering
  \caption{Percentage precision of parameters $\theta_{23}$ and $\delta_{CP}$ (as given in Eqn.~(\ref{prcision})) around the 
  true value for the synergistically combined data from NO$\nu$A + DUNE + UNO experiments, as depicted in Fig.~\ref{fig:totchicp}.}
  \begin{tabular}{|c|c|c c|c c||c|c|c c|c c|}
    \hline \hline
    \multicolumn{2}{|c|}{\textbf{True Value}} & \multicolumn{4}{c||}{\textbf{LO Precision}} 
                                                                     & \multicolumn{2}{c|}{\textbf{True Value}} & \multicolumn{4}{c|}{\textbf{HO Precision}}  \\
     \cline{1-12}
    \multirow{2}{*}{\textbf{{\bf \fontsize{15}{31}\selectfont $\theta_{23}$}}} 
    & \multirow{2}{*}{\textbf{{\bf \fontsize{15}{31}\selectfont $\delta_{CP}$}}} & \multicolumn{2}{c|}{\textbf{2$\sigma$}} & \multicolumn{2}{c||}{\textbf{3$\sigma$}}
    & \multirow{2}{*}{\textbf{{\bf \fontsize{15}{31}\selectfont $\theta_{23}$}}} 
    & \multirow{2}{*}{\textbf{{\bf \fontsize{15}{31}\selectfont $\delta_{CP}$}}} & \multicolumn{2}{c|}{\textbf{2$\sigma$}} & \multicolumn{2}{c|}{\textbf{3$\sigma$}}\\
    \cline{3-6} \cline{9-12}
   &  & \textbf{{\bf \fontsize{15}{31}\selectfont $\theta_{23}$}} 
    &  \textbf{{\bf \fontsize{15}{31}\selectfont $\delta_{CP}$}} & \textbf{{\bf \fontsize{15}{31}\selectfont $\theta_{23}$}} 
    &  \textbf{{\bf \fontsize{15}{31}\selectfont $\delta_{CP}$}} & & & \textbf{{\bf \fontsize{15}{31}\selectfont $\theta_{23}$}} 
    &  \textbf{{\bf \fontsize{15}{31}\selectfont $\delta_{CP}$}} & \textbf{{\bf \fontsize{15}{31}\selectfont $\theta_{23}$}} 
    &  \textbf{{\bf \fontsize{15}{31}\selectfont $\delta_{CP}$}} \\
         \cline{1-12} \cline{1-12} 
         \hline \hline
\multirow{3}{*}{\textbf{{\bf \fontsize{15}{31}\selectfont $38^0$}}}    & 90 & 0.92 & 5.00 & 2.11 & 17.50 
 & \multirow{3}{*}{\textbf{{\bf \fontsize{15}{31}\selectfont $49^0$}}} & 90 & 1.21 & 7.50 & 4.63 & 24.50 \\
 & 0   & 0.93 & 15.00 & 3.86 & 25.00 & & 0   & 2.25 & 13.75 & 4.96 & 26.25  \\
 & -90 & 0.92 & 7.50  & 2.62 & 20.00 & & -90 & 0.50 & 3.50  & 4.22 & 20.00  \\
 \cline{1-12}
 \multirow{3}{*}{\textbf{{\bf \fontsize{15}{31}\selectfont $41^0$}}}   & 90 & 2.21 & 7.50 & 12.11 & 25.00 
 & \multirow{3}{*}{\textbf{{\bf \fontsize{15}{31}\selectfont $52^0$}}} & 90 & 0.67 & 7.50 & 1.64  & 21.25 \\
 & 0   & 1.91 & 7.50 & 5.30  & 22.50 & & 0   & 0.87 & 15.25 & 1.83 & 25.00  \\
 & -90 & 2.10 & 5.00 & 10.09 & 22.50 & & -90 & 0.77 & 11.25 & 1.84 & 20.75   \\
   \hline
    \hline
  \end{tabular}
  \label{tab:precision}
\end{table}

In Table \ref{tab:precision}, we list the values of the $2\sigma$ and $3\sigma$ precision of $\theta_{23}$ and $\delta_{CP}$ 
using these expressions for the case of synergistically combined data of three experiments NO$\nu$A[10,0] + DUNE[10,0] + UNO[10,0].
We observe that the CP precision is seem to be better for $\delta_{CP} = \pm 90^0$ as compared to $\delta_{CP} = 0^0$
for a given true value of $\theta_{23}$. This is because at given $\theta_{23}$ for both NO$\nu$A (Fig.~\ref{fig:chicp8}) 
and DUNE (Fig.~\ref{fig:chicp13}) experiments, the solution 
around the true point with $\delta_{CP}=0^0$, either covers the whole $\delta_{CP}$ range or covers larger $\delta_{CP}$ range
in comparison to $\delta_{CP}$ range around the given true point with $\delta_{CP}= \pm 90^0$. 
This in turn means that for given $\theta_{23}$, the sensitivity towards the 
test $\delta_{CP}$ variations at $\delta_{CP}=0^0$ is more as compared 
to variations at $\delta_{CP}= \pm 90^0$.
But now if we take a look  at UNO 
experiment (Fig.~\ref{fig:chicp27}), 
sensitivity towards the $\delta_{CP}$ variations at 
$\delta_{CP}=0^0$ is almost same as at $\delta_{CP}= \pm 90^0$. 
The combined effect of all the experiments gives bit high 
sensitivity at $\delta_{CP}= 0^0$, thus accounts for bit worse precision. 
While the $\theta_{23}$ precision at given value of 
$\delta_{CP}$ is worse near
the maximal mixing and improves as one moves away.

%%%%%%%%%%%%%%%%%%%%%%%%%%%%%%%%%%%%%%%%%%%%%%%%%%%%%%%%%%%%%%%%%%%%%%%%%%%%%%%%%%%%%%%%%%%%%%%
\section{Conclusions and perspectives}
\label{section:Conclusion}
%%%%%%%%%%%%%%%%%%%%%%%%%%%%%%%%%%%%%%%%%%%%%%%%%%%%%%%%%%%%%%%%%%%%%%%%%%%%%%%%%%%%%%%%%%%%%%%
%  
The shape of the neutrino spectrum at the detector site feebly depends on the $\delta_{CP}$ phase, as 
is clear from Figs.~\ref{fig:10kty} and \ref{fig:5hkty}. It is also observable from these figures that,
the value of the event rate for detector configuration of 500 KTY is about 50 times 
that at 10 KTY detector configuration. We observe from Fig.~\ref{fig:entlp} that both 
NO$\nu$A and UNO experiments represent almost equal possibility to
investigate $\delta_{CP}$ phase.

From Figs.~\ref{fig:emu_dcp27}, \ref{fig:emu_dcp13} and \ref{fig:emu_dcp8}
we observe that $\mu^+$ mesons accelerated in to the energy range $30 \leqslant E_\mu \leqslant 50$ GeV 
exhibits observable sensitivity towards the $\delta_{CP}$ phase variations. Experimental setup
NO$\nu$A considers the highest precedence over the other two experiments, in respect of
having comparable sensitivity and highest mass ordering difference to signal ratio (i.e. $N_{asy}$)
value. The presence of internal degeneracy in the both upper and lower halves of $\delta_{CP}$ phase
in case of both NO$\nu$A and DUNE experiments hinders the investigation of narrow range of $\delta_{CP}$ phase.
But due to the absence of internal degeneracy for the UNO experiment, it provides the opportunity to 
investigate narrow ranges for $\delta_{CP}$ phase. It is advisable that to have highest sensitivity 
towards the $\delta_{CP}$ variations, we should choose $E_{\mu} \simeq 30$ GeV for UNO (L=2700 Km) and
$E_{\mu} \simeq 50$ GeV for DUNE (L=1300 Km) and NO$\nu$A (L=812 Km) experiments.

There is observable octant sensitivity in case of 500 KTY detector configurations, where as for the 
10 KTY configurations octant sensitivity is very low in comparison. Though UNO, 500 KTY experimental configuration 
alone provides observable octant sensitivity for both type of hierarchies, but the synergistic addition of 
three experiments data further enhances the sensitivity over both the test octants appreciably.

Though with the individual experimental data in respect of contour plots in the $\theta_{23} -\delta_{CP}$ plane, 
we come across various multiple discrete solutions like RO-W$\delta_{CP}$, WO-R$\delta_{CP}$ and WO-W$\delta_{CP}$ 
as well as continuous solutions arising due to submergence of different discrete solutions to true solution. 
But, synergistic addition of data from all the three considered experiments, removes all these discrete as well as 
continuous solutions up to 3$\sigma$ level very well, especially away from maximal mixing of atmospheric mixing
angle (i.e. $\theta_{23}= 45^0$).

We observe that for synergistically combined data of three experiments, the CP precision is seem to be better 
for $\delta_{CP} = \pm 90^0$ as compared to $\delta_{CP} = 0^0$ for a given true value of $\theta_{23}$. 
While the $\theta_{23}$ precision at given value of $\delta_{CP}$ is worse near the maximal mixing 
and improves as one moves away.
 %%%
%%%%%%%%%%%%%%%%%%%%%%%%%%%%%%%%%%%%%%%%%%%%%%%%%%%%%%%%%%%%%%%%%%%%%%%%%%%%%%%%%%%%%%%%%%%%%%%
\section*{Acknowledgement}
\label{section:acknow}
%%%%%%%%%%%%%%%%%%%%%%%%%%%%%%%%%%%%%%%%%%%%%%%%%%%%%%%%%%%%%%%%%%%%%%%%%%%%%%%%%%%%%%%%%%%%%%%
%  
I would like to thank Dr. Sanjib Kumar Agarwalla for useful discussions and providing data 
related to PREM density profile and line average of Earth's density.

%%\newpage
%%%%%%%%%%%%%%%%%%%%%%%%%%%%%%%%%%%%%%%%%%%%%%%%%%%%%%%%%%%%%%%%%%%%%%%%%               
%%%%%%%%%%%%%%%%%%%                           %%%%%%%%%%%%%%%%%%%%%%%%%%%

\end{document}